\documentclass[useAMS,usenatbib]{mn2e}
\usepackage{natbib}
\usepackage{graphicx}
\usepackage{amssymb}
\usepackage{epsfig}
\usepackage{float}
\usepackage{times}

\title[Luminous and dark matter mass ratio for massive galaxies to
$z\sim2$]{Clustering properties of galaxies selected in stellar mass:
  Breaking down the link between luminous and dark matter in massive
  galaxies from $\mathbf{z=0}$ to $\mathbf{z=2}$}
\author[S. Foucaud et al.]{S. Foucaud$^{1,2}$\thanks{E-mail:
    foucaud@ntnu.edu.tw}, C.~J.  Conselice$^{1}$, W.~G. Hartley$^{1}$, K.~P.
  Lane$^{1,3}$, S.~P. Bamford$^{1}$, \newauthor
  O. Almaini$^{1}$ and K. Bundy$^{4}$  \\
  $^{1}$School of Physics and Astronomy, University of Nottingham,
  University Park, Nottingham NG7 2RD\\
  $^{2}$Department of Earth Sciences, National Taiwan Normal University,
  N$^{\circ}$88, Tingzhou Road, Sec. 4, Taipei 11677, Taiwan (R.O.C.)\\
  $^{3}$Astrophysics, Department of Physics, University of Oxford,
  Denys Wilkinson Building, Keble Road, Oxford, OX1 3RH\\
  $^{4}$Department of Astronomy, University of California Berkeley,
  601 Campbell Hall, Berkeley, CA 94720-3411, USA\\ 
}

\def\solmm{$h^{-1}M_{\odot}\,$}

\def\solmg{$M_{\odot}\,$}
\def\mass{$10^{11}M_{\odot}\,$}
\def\hmass{$10^{11.5}M_{\odot}\,$}

\def\mpc{$h^{-1}$Mpc }

\def\plottwo#1#2{\centering \leavevmode
\epsfxsize=.45\textwidth \epsfbox{#1} \hfil
\epsfxsize=.45\textwidth \epsfbox{#2}}

\begin{document}

\date{Accepted 2010 March 12.  Received 2010 February 7; in original form 2009 February 13}

\pagerange{\pageref{firstpage}--\pageref{lastpage}} \pubyear{2010}

\maketitle

\label{firstpage}

\begin{abstract}
  We present a study on the clustering of a stellar mass selected
  sample of 18,482 galaxies with stellar masses $M_{*}>10^{10}$\solmg at
  redshifts $0.4<z<2.0$, taken from the Palomar Observatory Wide-field
  Infrared Survey.  We examine the clustering properties of these
  stellar mass selected samples as a function of redshift and stellar
  mass, and discuss the implications of measured clustering strengths in
  terms of their likely halo masses.  We find that galaxies with high
  stellar masses have a progressively higher clustering strength, and
  amplitude, than galaxies with lower stellar masses.  We also find
  that galaxies within a fixed stellar mass range have a higher
  clustering strength at higher redshifts. We furthermore use our
  measured clustering strengths, combined with models from
  \cite{MW02}, to determine the average total masses of the dark
  matter haloes hosting these galaxies.  We conclude that for all
  galaxies in our sample the stellar-mass-to-total-mass ratio is
  always lower than the universal baryonic mass fraction.  Using our
  results, and a compilation from the literature, we furthermore show
  that there is a strong correlation between
  stellar-mass-to-total-mass ratio and derived halo masses for
  central galaxies, such that more massive haloes contain a lower
  fraction of their mass in the form of stars over our entire redshift
  range. For central galaxies in haloes with masses $M_{\rm
    halo}>10^{13}$\solmm we find that this ratio is $<0.02$, much
  lower than the universal baryonic mass fraction. We show that the
  remaining baryonic mass is included partially in stars within satellite
  galaxies in these haloes, and as diffuse hot and warm gas. 
  We also find that, at a fixed stellar mass, the
  stellar-to-total-mass ratio increases at lower redshifts. This
  suggests that galaxies at a fixed stellar mass form later in lower
  mass dark matter haloes, and earlier in massive haloes.  We
  interpret this as a ``halo downsizing'' effect, however some of this
  evolution could be attributed to halo assembly bias.
\end{abstract}


\section{Introduction}
\label{sec:intro}

Astronomers in the last decade have made major progress in
understanding the properties and evolution of galaxies in the distant
Universe.   Galaxies up to redshifts $z\sim6$, and perhaps even
higher, have been discovered in large numbers allowing statistically
significant population characteristics to be derived
\citep[e.g.][]{2007ApJ...670..928B,2009MNRAS.395.2196M}.  We in fact now
have a good understanding of basic galaxy quantities, such as the
luminosity function
\citep[e.g.][]{2003ApJ...595..698D,2005A&A...439..863I,2007MNRAS.380..585C,ciras10},
as well as how scaling relations, such as the Tully-Fisher relation
evolve up to, at least, $z\sim 1.2$
\citep[e.g.][]{2004A&A...420...97B,2005ApJ...628..160C,2006MNRAS.366..308B,2006ApJ...653.1049W,2007ApJ...660L..35K,2007ApJ...668..846B,2007MNRAS.377..806C,2008A&A...484..173P,2009A&A...496..389F}.
We have also begun to trace the stellar mass evolution of galaxies, as
well as the star formation rate, determining when stars and stellar
mass was put into place in the modern galaxy population
\citep[e.g.][]{2006A&A...453..869B,2006ApJ...651..120B,2006A&A...459..745F,2009ApJ...701.1765M}.

Stellar masses are now becoming a standard measure for galaxies, and
are being used to trace the evolution of the galaxy population in
terms of star formation rates and morphologies
\citep[e.g.][]{2005ApJ...625..621B,2005ApJ...620..564C,2008MNRAS.383.1366C,2007ApJ...660L..47N,2007ApJ...663..834B,2008ApJ...686...72C}.
However, stellar masses only trace one aspect of the masses of
galaxies, and ideally and ultimately, we want to be able to measure
galaxy total masses, that include contributions from stars, gas, and
dark matter.  Galaxies are believed to be hosted by massive dark
matter haloes that make up more than 85\% of their total masses,
and thus clearly tracing the co-evolution of galaxies and their haloes
is a major and important goal.

Measuring dynamical or total masses for galaxies is, however, very
difficult, as it requires observations that are very challenging to
obtain, or requires unusual and rare circumstances such as
gravitational lensing.  For example, dynamical masses can be measured
through rotation curves, or internal velocity dispersions, but this
becomes difficult at higher redshifts, with reliable internal
velocities existing for only a handful of galaxies at $z\sim1.0$ and
above
\citep[e.g.][]{2005ApJ...628..160C,2006ApJ...645.1062F,2006MNRAS.368.1631S,2007ApJ...668..846B,2008A&A...488...99V}.
Furthermore, it is difficult to know whether these kinematic measures
are tracing the total potential, or just the inner parts of galaxies.
Likewise, using gravitational lensing to measure galaxy total masses
is difficult, as it requires lensed background galaxies, and such
instances are very rare. It is also not yet clear if these galaxies
are representative, as they may have special mass profiles that are
conducive to lensing. The weak galaxy-galaxy lensing technique
provides another way to estimate the total masses of galaxies. No
unusual circumstances are required, but individual measurements are
extremely difficult to achieve. Stacking techniques give reliable
results, but involve combining galaxies in haloes of different masses,
which complicates the interpretation
\cite[e.g.][]{2006MNRAS.368..715M}. Stacked satellite kinematics can
also be used to probe the total masses of galaxies, but face the same
problems as the weak galaxy-galaxy lensing method
\citep{2009MNRAS.392..917M}. Therefore, obtaining total mass estimates
for galaxies is difficult, and very few measures, or even estimates,
have been produced for galaxies outside the local Universe.

One very powerful method for measuring the total masses of galaxies is
to measure their clustering.  Clustering measurements are independent
of photometric properties, and as such they can be used to highlight
fundamental properties of galaxy populations without assumptions
concerning stellar populations or mass profiles.  Previously, it has
been shown that galaxies with higher stellar masses cluster more than
systems with lower stellar mass, with a very strong clustering above the
characteristic stellar mass $M^{*}$ of the Schechter mass function
\citep{2006MNRAS.368...21L}.  In the halo model of galaxy formation,
the large-scale distribution of galaxies is determined by the
distribution of dark matter haloes \citep{MW02}. Therefore, halo
clustering is a strong function of halo mass, with more massive haloes
more strongly clustered, providing a means to study the relationship
between galaxy properties, and dark matter halo masses. In
fact, basic calculations allow one to convert a clustering strength
for a sample of galaxies into a corresponding halo mass in which
galaxies reside
\citep[e.g.][]{2004ApJ...611..685O,2008MNRAS.383.1131M,2008ApJ...679..269Y}.

Clustering measurements have been performed on galaxy samples in both
the nearby and distant Universe.  In the local Universe, studies have
established that clustering depends on the type of galaxy
under consideration; for example, early-type, red, galaxies are more
clustered than late-type, blue galaxies
\citep[e.g.][]{1997ApJ...489...37G,2002MNRAS.332..827N,2007MNRAS.379.1562C,2009MNRAS.392..682C},
and luminous galaxies are more clustered than faint galaxies
\citep[e.g.][]{2001MNRAS.328...64N,
  2005ApJ...630....1Z,2006MNRAS.369...68S}.  Moreover, distant massive
galaxies selected by their extremely red colours have been shown to
strongly cluster, and therefore are thought
to inhabit massive dark matter haloes
\citep[e.g.][]{UDS-EDR-DRG,2007ApJ...654..138Q,2008MNRAS.383.1131M,hartley08}.

In this paper, we present the first general study of the clustering
properties for a stellar mass selected sample of galaxies up to
$z\sim2$.  We carry this out by measuring the correlation length and amplitude
for galaxies selected with stellar masses $M_*>10^{10}$\solmg within
the Palomar Observatory Wide-field Infrared Survey (POWIR)
\citep{2006ApJ...651..120B,2008MNRAS.383.1366C}, which covers the
fields where spectroscopy and other multi-wavelength data are available
through the Deep Extragalactic Evolutionary Probe 2 (DEEP2) survey and
All-wavelength Extended Groth strip International Survey (AEGIS)
\citep{2003SPIE.4834..161D,2007ApJ...660L...1D}.  In total we examine
the clustering strength for 18,482 galaxies selected by stellar mass
within 0.7 deg$^{2}$. We derive an estimate of the total masses for
these galaxies, and study in detail their stellar-mass-to-total-mass
ratio.

This paper is presented as follows: our data-sets, catalogues and our
photometric redshifts and stellar mass estimations are described in
Section~\ref{sec:data}; in Section~\ref{sec:cf} we describe the
methods used in measuring the clustering properties of our samples;
Section~\ref{sec:dmm} is dedicated to the methods we use to derive the
masses of dark matter haloes for our samples; in
Section~\ref{sec:ratio} we compare our results with the literature and
models; and finally we summerise our conclusions in
Section~\ref{sec:summ}. Throughout the paper, we assume a $\Lambda$CDM
cosmology with $\Omega_m=0.3$, $\Omega_{\Lambda}=0.7$,
$h=H_0/70$~km~s$^{-1}$~Mpc$^{-1}$. To ease comparisons with previous
work, we use a concordance model with fiducial values of $n_s=1.0$ and
$\sigma_8=0.9$. To determine stellar masses throughout this paper, we
use the Initial Mass Function (IMF) from \cite{chabrier03} and assume
a Hubble constant of $H_0=70$~km~s$^{-1}$~Mpc$^{-1}$.

\section{The Palomar/DEEP2 survey}
\label{sec:data}

\subsection{Data sets}
\label{sec:data-gen}

All of the galaxies in this paper are found within three of the four
fields covered by the Palomar Observatory Wide-Field Infrared Survey
\citep[POWIR, Table~1;][]{conselice_mass}.  The POWIR survey was
designed to obtain deep $K$-band and $J$-band data over a significant
area ($\sim$1.5 deg$^2$).  Observations were carried out between
September 2002 and October 2005 over a total of $\sim 70$ nights. This
survey covers the Extended Groth Strip \citep{2007ApJ...660L...1D},
and three other fields that the DEEP2 team has observed with the
DEIMOS spectrograph \citep{2003SPIE.4834..161D}.  The total area imaged in
the $K$-band is 4920 arcmin$^{2}$=1.37~deg$^{2}$, with half of this
area imaged in the $J$-band. The goal depth was $K_{\rm s,vega}=21$,
although not all fields are covered to this depth, therefore we
select the fields which have 5$\sigma$ depths between 
$K_{\rm s,vega}=20.2-21.5$ for point sources, measured in a 2\arcsec\,
diameter aperture.  For our purposes we abbreviate the fields covered
as: EGS (Extended Groth Strip), Field 2, Field 3, and Field 4
(Table~\ref{tab1}). In the following study we use data in the EGS
field only to train our photometric redshifts (see
section~\ref{sec:data-mass-z}), while only the other three fields are
used to perform our clustering analysis (see
section~\ref{sec:cf-acf}). For extensive information on this survey,
and the data products we use from it, see \cite{2006ApJ...651..120B},
\cite{conselice_mass} and \cite{2008MNRAS.383.1366C}.

All of the $K_{\rm s}$-band data were acquired utilising the Wide
Field Infrared Camera (WIRC) on the Palomar 5 meter telescope.
WIRC has an effective field of view of $8.1\arcmin \times 8.1\arcmin$,
with a pixel scale of 0.25\arcsec pixel$^{-1}$.  The total survey
contains 75 WIRC pointings.  During the $K_{\rm s}$-band observations
we used 30 second integrations, with four exposures per pointing.  The
$J$-band observations were taken with 120 second exposures per
pointing.  Typical total exposure times were between one and two hours
for both bands. The reduction procedure follows a standard method
for combining near-infrared (NIR) ground-based imaging, and is
described in more detail in \cite{2006ApJ...651..120B}. The resulting
seeing FWHM in the $K_{\rm s}$-band imaging ranges from 0.8'' to
1.2'', and is typically 1.0''.  Photometric calibration was carried
out by referencing Persson standard stars during photometric
conditions, which were later cross calibrated with 2MASS stars
\citep{2mass}.

\setcounter{table}{0}
\begin{table}
 \begin{tabular}{@{}lcccc}
  \hline
Field & RA & Dec. & K-band area & J-band area \\
 & & & (arcmin$^{2}$) & (arcmin$^{2}$) \\
\hline
{\it EGS} & 14 17 00 & +52 30 00 & 2165 & 656 \\
{\bf Field 2} & 16 52 00 & +34 55 00 & 787 & 0 \\
{\bf Field 3} & 23 30 00 & +00 00 00 & 984 & 984 \\
{\bf Field 4} & 02 30 00 & +00 00 00 & 984 & 787 \\
\hline
 \end{tabular}
 \caption{The Palomar Fields and WIRC pointings areas. The EGS field is used 
         in this study for training the photometric redshifts, and the other 
         three for measuring clustering properties.}
 \label{tab1}
\end{table}

The final NIR images were made by combining individual mosaics
obtained over different nights. The $K_{\rm s}$-band mosaics are
comprised of coadditions of $4 \times 30$ seconds exposures dithered
over a non-repeating 7.0'' pattern.  The images were processed using a
double-pass reduction pipeline developed specifically for WIRC. The
detection and photometry of our galaxies was performed using the SExtractor
package \citep{sextractor}. False artifacts are removed through
SExtractor flags which identify sources that do not have normal galaxy
or stellar profiles.  From this we built a $K$-selected sample and
then cross-referenced these sources with the DEEP2 redshift catalogue.

Optical imaging from the Canada-France-Hawaii Telescope (CFHT), over
all fields, is used to estimate photometric redshifts and stellar
masses, with the help of spectroscopy from the DEIMOS spectrograph on
the Keck II telescope \citep{2003SPIE.4841.1657F}.  This optical
imaging comes from the CFHT 3.6-m, and consists of data in the $B$-,
$R$- and $I$-bands taken with the CFH12K camera - a 12,288 $\times$
8,192 pixels CCD mosaic with a pixel scale of 0.21\arcsec.  The
integration times for these observations are 1 hour in $B$ and $R$,
and 2 hours in $I$, per pointing, with a $R$-band 5$\sigma$ depth of
$R_{\rm AB}=25.1$, and similar depths in $B$ and $I$
\citep{2004ApJ...617..765C,conselice_mass,2008MNRAS.383.1366C}.  From this imaging data a
$R_{\rm AB}=24.1$ magnitude limit was used for determining targets
for the DEEP2 spectroscopy.  The seeing for the optical imaging is
roughly the same as that for the NIR imaging, and we measure
photometry consistently, using a 2\arcsec diameter aperture.
  
The Keck spectra were acquired with the DEIMOS spectrograph as part of
the DEEP2 redshift survey \citep{2003SPIE.4834..161D}.  The selection
of targets for the DEEP2 spectroscopy was based on the optical
properties of the galaxies detected in the CFHT photometry, with the
basic selection criteria $R_{\rm AB}<24.1$.  Objects in Fields 2, 3
and 4 were selected for spectroscopy based on their position in
$(B-R)$ vs.  $(R-I)$ colour space to focus on galaxies at redshifts
$z>0.7$.  Spectroscopy in the EGS was acquired using the magnitude
limit, but to avoid the survey being completely dominated by lower
redshift galaxies, sources with colours suggesting they are at $z<0.7$
were down-weighted in favour of sources with $z>0.7$. The total survey
targeted over 30,000 galaxies, with about a third of these in the EGS
field.  In all fields the sampling rate for galaxies that meet the
selection criteria is 60\%.  The DEIMOS spectroscopy was obtained
using the 1200 line/mm grating, with a resolution R $\sim5000$
covering the wavelength range 6500 - 9100 \AA.  Redshifts were
measured by comparing templates to data, and we only utilise those
redshifts determined by the identification of two or more lines,
providing very secure measurements.

Since accurate photometry, and photometric errors are important for
the purposes of this paper we therefore give some details about our
methods of measuring magnitudes.  The photometric errors, and the
detection limit of each $K$-band image, are estimated by randomly
inserting simulated galaxies of known magnitude, surface brightness
profile, and size into each image, and then recovering these simulated
objects with the same detection parameters used for real objects
\citep[see][]{2006ApJ...651..120B,2008MNRAS.383.1366C}.  To determine
the detection and photometric fidelity of the images in more detail,
two sets of 10,000 mock galaxies were created, each with an intrinsic
exponential and de Vaucouleurs profile \citep[see][]{conselice_mass}.

Systematic errors in the measurement of magnitudes, due to the
detection method, were also estimated using the same simulations.  The
recovery fraction is found to be essentially 100\% at the magnitudes
of our sample galaxies.  This was determined through simulating
galaxies down to $K = 21$, as detailed in \cite{conselice_mass}.
However, all of the galaxies in our sample are at least a
magnitude brighter
than this, with all galaxies within our stellar mass limits having
magnitudes $K < 20$ as shown in \cite{conselice_mass}.
\cite{2007MNRAS.382..109T} and \cite{conselice_mass} furthermore
carried out extensive simulations to show that at this limit we are
nearly 100\% complete and are retrieving nearly all of the light from
these galaxies.  However, depending on the sizes of the galaxies, as
much as 0.2 mag in the recovered $K$-band light could be missed,
although for any single galaxy it is unlikely that this amount of
light is missed due to the surface brightness profiles of our objects
\citep{2007MNRAS.382..109T}.  This issue is addressed in great detail
in \cite{conselice_mass}, but in summary we account for this
uncertainty in the error budget.

\subsection{Redshifts and Stellar Masses}
\label{sec:data-mass}

\subsubsection{Photometric Redshifts}
\label{sec:data-mass-z}

We calculate photometric redshifts for our $K$-selected galaxies which
do not have DEEP2 spectroscopy.  These photometric redshifts are based
on the optical+near infrared imaging, in the $BRIJK$ (or $BRIK$ for
half the sample) bands, and are fit in two ways, depending on the
brightness of a galaxy in the optical.  For galaxies that meet the
spectroscopic criteria, $R_{\rm AB}<24.1$, we utilise a neural network
photometric redshift technique to take advantage of the large number
of secure redshifts with similar photometric data.  Most of the
$R_{\rm AB}<24.1$ sources not targeted for spectroscopy should be
within $z<1.4$, as very few objects this bright are found at higher
redshifts \citep[e.g.][]{2004ApJ...604..534S}. The neural network
fitting is done through the use of the ANNz
\citep{2004PASP..116..345C} method and code.  To train the code, we
use the $\sim5000$ redshifts in the EGS field, whose galaxies are
selected with a magnitude limit of $R_{\rm AB}<24.1$ and span
the redshift range $0.4 < z < 1.4$ (see section~\ref{sec:data-gen}). We then
use this training to calculate the photometric redshifts for galaxies
with $R_{\rm AB}<24.1$ in all fields.  The overall agreement between
our photometric redshifts and our ANNz spectroscopic redshifts is very
good using this technique, with $\delta z/(1+z)=0.07$ out to
$z\sim1.4$. The agreement is even better for the $M_{*}>$\mass
galaxies where we find $\delta z/(1+z)=0.025$ across all of our four
fields.  The photometry we use for our photometric redshift estimates
are measured within a 2\arcsec\, diameter aperture.

For galaxies fainter than $R_{\rm AB}=24.1$ we estimate photometric
redshifts using Bayesian techniques based on the software from
\cite{2000ApJ...536..571B}.  For an object to have a photometric
redshift we require that it be detected at the 5$\sigma$ level in all
optical and near-infrared bands ($BRIJK$), which in the $R$-band
reaches $R_{\rm AB}=25.1$ \citep{2004ApJ...617..765C,conselice_mass,2008MNRAS.383.1366C}.
We optimise our results, and correct for systematics, through a comparison with
spectroscopic redshifts, resulting in a redshift accuracy of $\delta
z/(1+z)=0.17$ for $R_{\rm AB}>24.1$ systems based on comparisons to
redshifts from \cite{2006ApJ...653.1004R}.  These $R_{\rm AB}>24.1$
galaxies are however only a very small part of our sample. Up to
$z\sim1.4$ only 6 (2.6\%) of our $M_{*}>$\hmass galaxies are in this
regime, while 412 (9\%) of our \mass$<M_{*}<$\hmass galaxies have an
$R$-band magnitude this faint, with a similar fraction for galaxies
down to M$_{*} = 10^{10}$ \solmg.  Furthermore, all of these systems are
at $z>1$.  At $z>1.4$ all of our sample galaxies are measured through
the \cite{2000ApJ...536..571B} method due to a lack of training
redshifts.  We also compared the results at low redshifts
between the \cite{2004PASP..116..345C} and \cite{2000ApJ...536..571B}  methods and their agreement is very good, at
the same level than the comparison with spectroscopic redshifts.

A thorough explanation, including error budgets and
uncertainties, for our redshifts are included in \cite{conselice_mass}
and \cite{2008MNRAS.383.1366C}.

\subsubsection{Stellar Masses}
\label{sec:data-mass-mass}

We match our $K$-band selected catalogues to the CFHT optical data to
obtain spectral energy distributions (SEDs) for all of our sources,
resulting in measured $BRIJK$ magnitudes.  From these we compute stellar
masses based on the methods and results outlined in
\cite{2005ApJ...625..621B}, \cite{2006ApJ...651..120B} and
\cite{conselice_mass}.

The basic mass fitting method consists of fitting a grid of model SEDs
constructed from \cite{BC03} (BC03) stellar population synthesis
models, with different star formation histories. We use
exponentially declining models to characterise the star formation
history, with various ages, metallicities and dust contents included.
These models are parameterised by an age, and an e-folding time for
parameterising the star formation history, where SFR $\alpha\,
e^{\frac{t}{\tau}}$.  The values of $\tau$ are randomly selected from
a range between 0.01 and 10 Gyr, while the age of the onset of star
formation ranges from 0 to 10 Gyr. The metallicity ranges from 0.0001
to 0.05 (BC03), and the dust content is parametrised by $\tau_{\rm
  V}$, the effective V-band optical depth for which we use values
$\tau_{\rm V}=0.0,0.5,1,2$.  Although we vary several parameters,
the resulting stellar masses from our fits do not depend strongly on
the various selection criteria used to characterise the age and the
metallicity of the stellar population.

We match magnitudes derived from these model star formation histories
to the actual data to obtain a measurement of stellar mass using a
Bayesian approach.  We calculate the likely stellar mass, age, and
absolute magnitudes for each galaxy at all star formation histories,
and determine stellar masses based on this distribution.
Distributions with larger ranges of stellar masses have larger
resulting uncertainties.  Typical errors for our stellar masses are
0.2 dex from the width of the probability distributions.  There are
also uncertainties from the choice of the Initial Mass Function (IMF).
Our stellar masses utilise the \cite{chabrier03} IMF, which can be
converted to Salpeter IMF stellar masses by dividing by a factor $\sim
1.5$ \citep[see][]{chabrier03}.  There are additional random
uncertainties due to photometric errors.  The resulting stellar masses
thus have a total random error of 0.2-0.3 dex, roughly a factor of
two.  The details behind these mass measurements and their
uncertainties, including the problem of thermal-pulsating AGB stars,
is described in \cite{2006ApJ...651..120B} and \cite{conselice_mass}.
We examine the changes in our stellar masses by using Bruzual \& Charlot
models updated in 2007 to include 
TP-AGB stars, where we find at most a decrease of 0.07 dex, or stellar masses
which are 20\% less than what we calculate using models without
TP-AGB stars.
Furthermore, \cite{2009ApJ...699..486C} have recently shown that
stellar mass estimates, from stellar population synthesis, suffer
systematic uncertainties mainly due to key phases of stellar evolution
and the initial mass function. However we are consistent in our
analysis as we are comparing results from different studies based on the same
stellar population models to determine stellar masses.
Throughout this work, stellar mass is given in unit of $M_{\odot}$,
having been computed assuming a Hubble constant of
$H_0=70$~km~s$^{-1}$~Mpc$^{-1}$.

In this study we only examine massive galaxies with stellar masses
$M_{*}>10^{10}$\solmg.  In fact, we examine the clustering
properties of galaxies within the following four mass selected samples
$10^{10.0}$\solmg$<M_*\leq10^{10.5}$\solmg,
$10^{10.5}$\solmg$<M_*\leq10^{11.0}$\solmg,
$10^{11.0}$\solmg$<M_*\leq10^{11.5}$\solmg and
$10^{11.0}$\solmg$<M_*\leq10^{12.0}$\solmg.  We also examine these
stellar mass cuts within four
different redshift bins: $0.4<z\leq0.8$, $0.8<z\leq1.2$, $1.2<z\leq1.6$
and $1.6<z\leq2.0$.  As discussed in \cite{conselice_mass}, our
lower-mass samples are largely incomplete in the higher redshift bins,
and the volume we are sampling at lower redshift does not allow us to
examine a large enough sample of massive galaxies. The abundances of
our different samples are reported in Table~\ref{tab:results1}.  We
also note that our final two stellar-mass bins are not independent, as
the third one overlaps completely with the fourth one. We decided to
keep the bins this way for statistical reasons, as otherwise it would
have been impossible to conduct studies on an independent higher stellar mass
bin within the highest redshift bins.  As we will see later, the
results for our two higher mass bins are very similar, and this
division does not effect our analysis.

\subsubsection{Photometric Redshift Errors}
\label{sec:data-mass-err}

In order to better constrain the errors introduced in our clustering
and abundance analysis due to possible inaccurate photometric 
redshifts, we built a set of five supplementary simulated catalogues based on
a series of Monte Carlo simulations.  We perform this simulation by
altering the redshifts of each of the galaxies in our sample, randomly
within our well calibrated and known photometric redshift errors, as based on
comparisons to spectroscopic redshifts.  The new stellar masses are then 
estimated with these altered redshifts for each object.

After recalculating these masses from our simulated redshifts,
we then re-select the galaxies within our different redshift and stellar
masses bins as in our original analysis, but by using these new
simulated galaxy catalogues.  The abundances of our different samples 
reported in Table~\ref{tab:results1}, are an average over all the
catalogues (the original one and the five altered ones), and an average 
over our three fields. The error budget listed in this table takes into 
account the errors introduced by the photometric redshift, and the 
resulting stellar mass errors, as accounted for by the rms 
over the six catalogues, plus the errors introduced by the cosmic variance as
measured from the field-to-field variance over our three fields.  Within
this paper we systematically estimate our errors by accounting for
these two effects. This is likely an over-estimation of the total error,
as the error on the photometric redshifts are partially included in the
field-to-field variance.

We also use these catalogues to estimate the fraction of interlopers
that are contaminating each of our redshift bins. As the
determination of stellar mass is directly linked with the
photometric redshift determination, it is very difficult to conduct an
interloper analysis for both selections at the same time. We
therefore focus on the interlopers between the different redshift
bins, which is likely to have the highest impact on our clustering
analyses.  Furthermore, our measurements are made on three independent
fields, all of which give similar results.  \cite{conselice_mass} also
performed a Monte-Carlo simulation in order to quantify the effect of
stellar mass errors on the density and mass function of our samples.
They found that the systematic effect is minor compared to the effect
of the cosmic variance, which we also find, and for which we account for
as well.

The interloper fraction within the different redshift bins is determined
by identifying objects that are within the same redshift bin in all
our simulated catalogues. Then the number of potential interlopers is 
identified in each catalogue based on the difference between this standard 
set and the additional number of galaxies seen in each. However, this method of
accounting for the photometric redshift errors increases 
artificially the number of interlopers, from the most populated redshift 
bins to the least populated. By comparing our five simulated catalogues with 
our original one, we find that this effects account for $\sim 10\%$ of the galaxies 
in the simulated catalogues. Taking into account this 
effect, we estimate that the maximum possible interloper fraction is: for
$0.4<z<0.8$, $f=21.6\pm1.8\%$, for $0.8<z<1.2$, $f=19.8\pm1.7\%$, for
$1.2<z<1.6$, $f=21.2\pm2.0\%$, and for $1.6<z<2.0$, $f=29.2\pm5.2\%$.

Our interloper fraction is therefore at most between 20\% and 30\% depending on
the redshift bin. Given the errors on our photometric redshift
determination, as quoted in Section~\ref{sec:data-mass-z}, this is of the
order we expect.   Furthermore, we note that this fraction of interlopers
is largely due to galaxies near the boundaries of our strict stellar mass and
redshift cuts entering other bins.  These galaxies typically have just 
slightly different redshifts and stellar masses before they were 
simulated, creating a slightly different population within each simulated 
redshift bin which is not significantly different from the original one.  
Regardless, we fully account for this effect within our measured clustering 
strengths error budget.

\begin{figure*}
\begin{center}
  \resizebox{0.75\hsize}{!}{\includegraphics{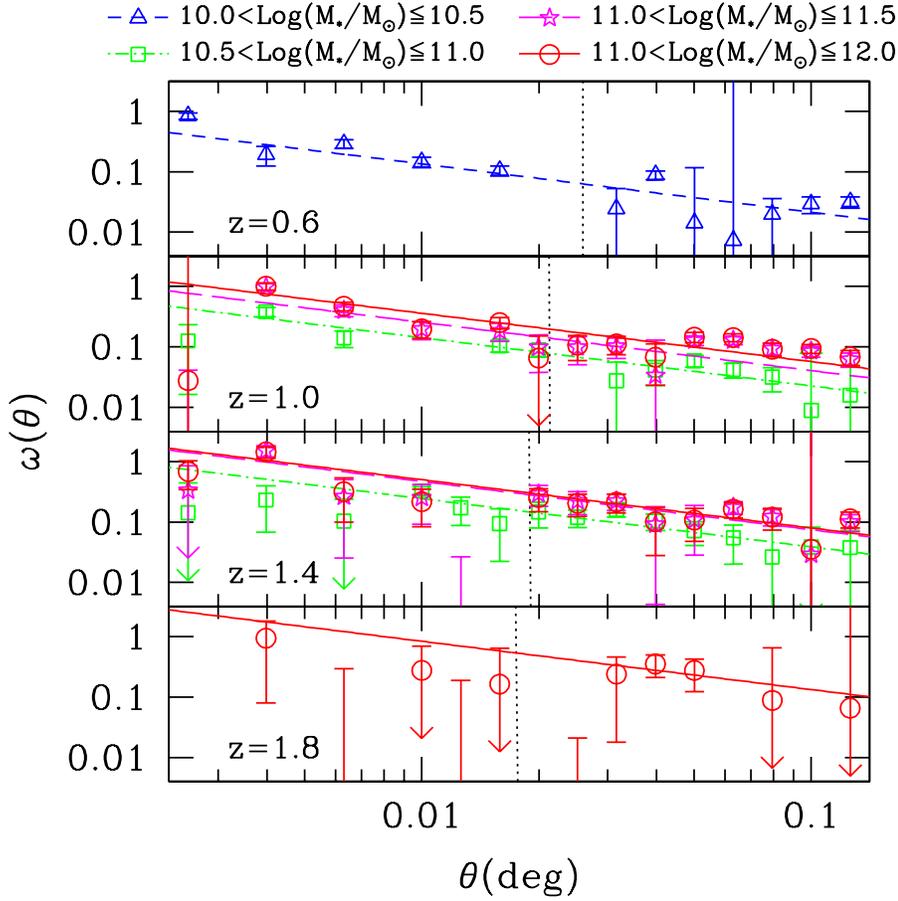}}
  \vspace{-0.5cm}
  \caption{The 2-point angular correlation function as a function of redshift
    for our mass-selected samples in our field 3 (see text). The open
    triangle, open square, open star and open circle symbols represent
    the measurements made for the mass selected samples at stellar
    masses $10^{10.0}$\solmg$<M_*\leq10^{10.5}$\solmg,
    $10^{10.5}$\solmg$<M_*\leq10^{11.0}$\solmg,
    $10^{11.0}$\solmg$<M_*\leq10^{11.5}$\solmg and
    $10^{11.0}$\solmg$<M_*\leq10^{12.0}$\solmg, respectively. The best
    fitted lines for the correlation function are represented by the
    short-dashed, dashed-dotted long-dashed and continuous lines,
    respectively for the different mass samples as described above.
    The vertical dotted line corresponds to the lower limit at
    $\sim1$\mpc of the range over which our data are fitted, in order to avoid any
    excess of pairs due to multiple halo occupation.}
  \label{fig:cf}
\end{center}
\end{figure*}

\section{Clustering properties of massive galaxies}
\label{sec:cf}

\subsection{Angular clustering}
\label{sec:cf-acf}

The primary goal of this work is to study the clustering properties of
galaxies selected according to their stellar mass. A clustering
analysis is a study of the distribution of galaxies at all scales.
Given its peculiar geometry, the EGS field is not ideally suited for
such a study. Indeed, as galaxies are distributed along a strip in the
EGS, the clustering properties at the largest scales are difficult to
establish given the small numbers of objects we have in each stellar
mass and redshift bin in this field. For the other three fields we
mask out overlap regions, and other regions which do not have
reliable photometry. This masking of regions empty of galaxies
is then taken into account in the following analysis. Overall we 
performed our analyses over a total area of 0.7 deg$^2$.

Within our clustering analysis we first measure the 2-point angular
correlation function $\omega(\theta)$ for our sample using the
\cite{LS} estimator:

\begin{equation}
\omega ( \theta) ={\mbox{DD} - 2\mbox{DR} + \mbox{RR}\over \mbox{RR}}
\label{eq:LS}
\end{equation}

\noindent where the DD, DR and RR terms refer to the number of data-data,
data-random and random-random galaxy pairs having angular separations
between $\theta$ and $\theta+\delta\theta$.
Figure~\ref{fig:cf} shows the correlation function
derived from our mass-selected samples in different redshift bins for
one of our fields.  

The best fit for the angular correlation is assumed to be a power-law
of the form \citep{1977ApJ...217..385G}:

\begin{equation}
\omega(\theta)=A_{\omega}(\theta^{-\delta}-C_{\delta})
\label{eq:wtheta}
\end{equation}

\noindent with $A_{\omega}$ the amplitude at 1 degree, $\delta$ the slope, 
and $C_{\delta}$ the integral constraint due to the limited area of the
survey. For statistical reasons we fix the slope to the commonly used
fiducial value of $\delta=0.8$ \citep{1977ApJ...217..385G}.  We also
estimate the integral constraint ($C_{\delta}$) as follows
\citep{RSMF},

\begin{equation}
C_{\delta} = {1 \over {\Omega^2}} \int \! \int \theta^{-\delta} d\Omega_1 d\Omega_2
\label{eq:intcst}
\end{equation}

\noindent where $\Omega$ is the area subtended by the survey field. To 
determine $C_{\delta}$ we numerically integrate this expression over
each field, excluding masked regions. Assuming a slope of
$\delta=0.8$, the unmasked areas ($S_F$) covered by the fields, and
the corresponding integral constraint values are: $S_{F2}=639.55$
arcmin$^2$ and $C_{0.8\,F2}=3.61$, $S_{F3}=777.30$ arcmin$^2$ and
$C_{0.8\,F3}=3.76$, $S_{F4}=731.04$ arcmin$^2$ and $C_{0.8\,F4}=3.69$,
for fields 2, 3 and 4 respectively.

In order to fit the clustering reliably, we avoid the small-scale
excesses due to possible multiple galaxy occupation of a single dark
matter halo, as has been shown to exist in recent studies
\citep[e.g.][]{2005ApJ...635L.117O}. The lower bound on which we fit
our correlation was chosen to correspond to $\sim1$\mpc, the
observed one-halo term limit established by
\cite{2004ApJ...608...16Z}. These lower bounds are shown by the
vertical dotted lines on Figure~\ref{fig:cf}, and it is worth noting
that no apparent excess is displayed at the smallest scales.

The error-bars shown on Figure~\ref{fig:cf} are estimated from a
jackknife Monte-Carlo method. The error estimation for each scale is
made following the bootstrap method. In each sub-sample, objects are
randomly removed and duplicated from our position catalogue, and the
two-point correlation function is measured again. The value of the
correlation function at each scale is estimated from the mean of these
bootstrap catalogues, and errors are derived from the variance.
We then derive the best fit amplitude for each of our fields at 1
degree using a Marquardt least-linear method, that take into account
our bootstrap errors, and provide an estimate of the error on
the fitting.  We then derive a mean value from our six catalogues and over
each of our three fields. Table~\ref{tab:results1} summarises the values
measured for our samples.

As explained in Section~\ref{sec:data-mass-err}, the variance over the
six catalogues is used as an estimate for the systematic errors in the
photometric redshift determination and the variance over the three
field measurements is used as an estimate of the cosmic variance of
our samples. Our total error budget is a quadratic sum of the
fitting errors, photometric redshift errors, and cosmic variance errors.
As discussed in \cite{conselice_mass}, using models from
\cite{2004ApJ...600L.171S}, the cosmic variance on the number counts
in the redshift range we use ($0.4<z\leq2.0$) range from
$\sigma_v\sim0.1-0.2$ for our less massive samples, up to
$\sigma_v\sim0.3-0.5$ for our most massive ones. This is comparable
with the uncertainties we derived based on our field-to-field
measurements on the clustering analysis.

\begin{figure*}
 \begin{center}
 \plottwo{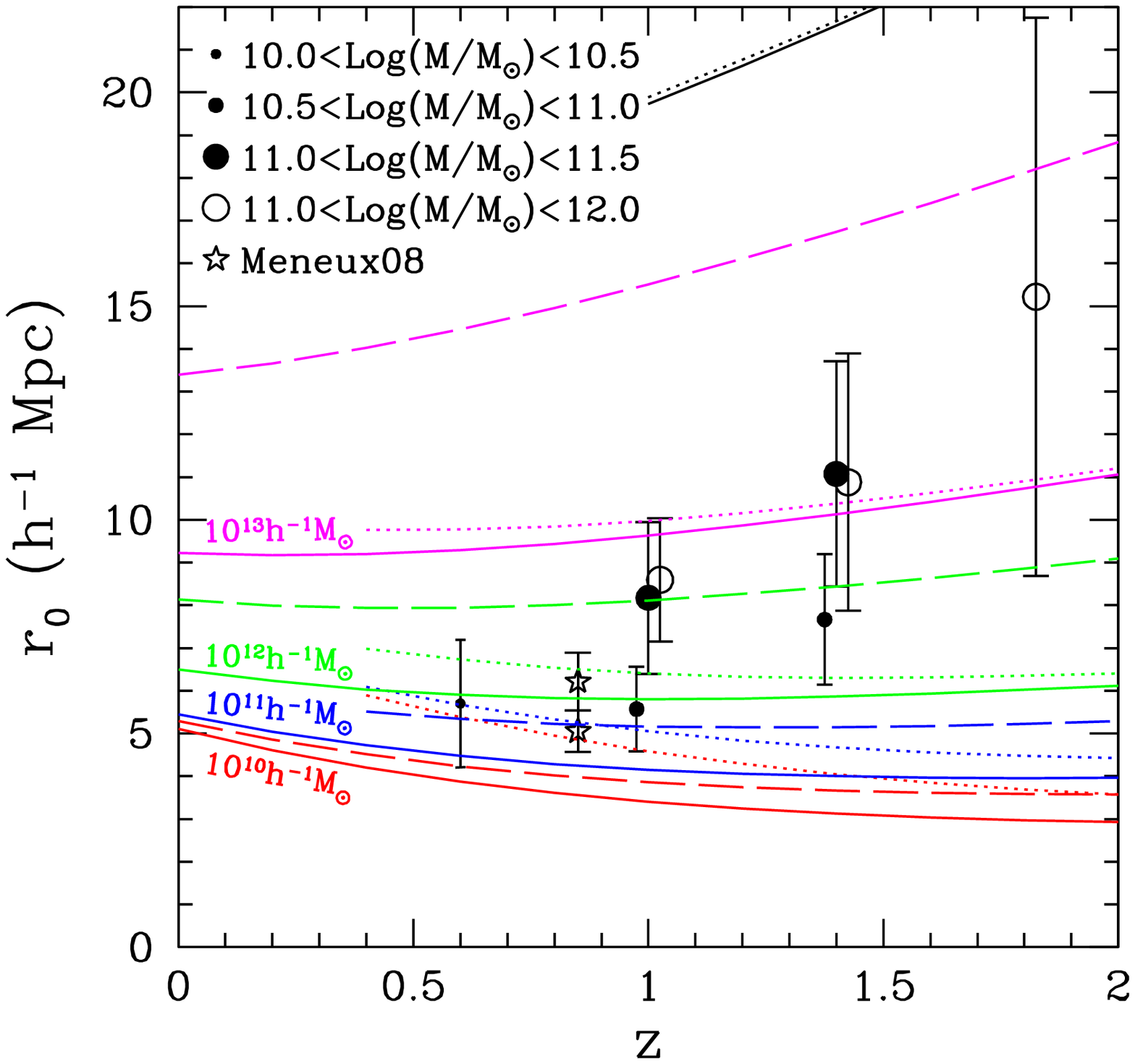}{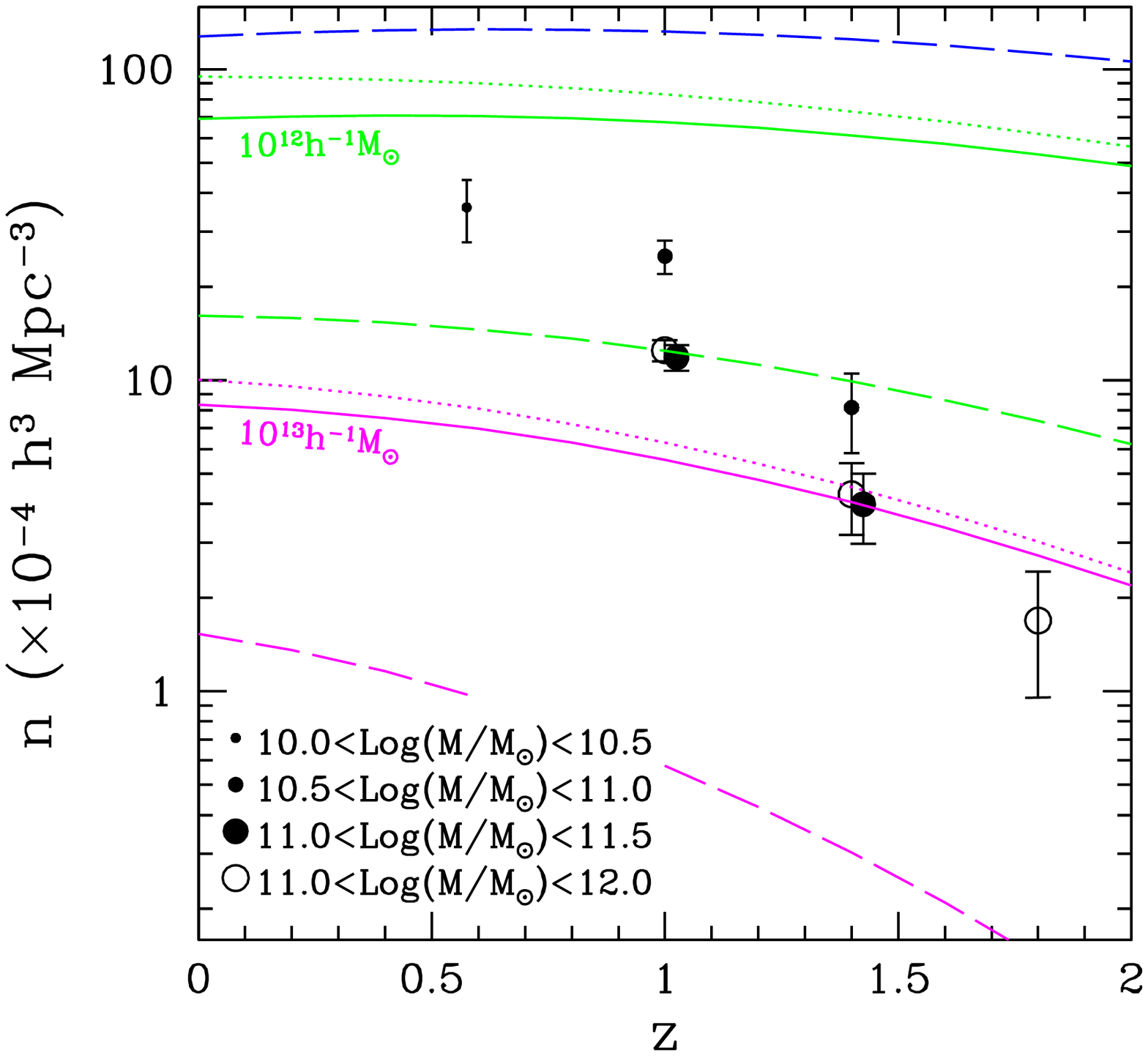}
  \vspace{-0.5cm}
 \caption[The measured correlation length and galaxy abundance as a function of redshift]{
   {\bf (a)} Correlation length ($r_0$) as a function of redshift.
   The correlation lengths measured for our different samples are
   represented by the full circles with the size of the circle
   increasing with the stellar mass of the samples. We also show
   measurements performed by \cite{2008A&A...478..299M} for two mass
   ranges (Log~$[M_*/(h_{100}^{-2}M_{\odot})]=10.22$ and $10.70$). The
   lines represent predictions from models by \cite{MW02} for
   different minimum masses of dark matter haloes (plain lines for
   $M_{\rm min}=10^{11.0}$\solmm, $10^{12.0}$\solmm and
   $10^{13.0}$\solmm, and dashed lines for $M_{\rm
     min}=5\times10^{11.0}$\solmm, $5\times10^{12.0}$\solmm and
   $5\times10^{13.0}$\solmm) while the dotted lines represent the
   predictions from models including halo distributions as described
   in \cite{2007ApJ...667..760Z}. {\bf (b)} Abundance ($n$) as a
   function of redshift.  The abundances measured for our different
   samples are represented by the full circles with the size of the
   circle increasing with the mass of the sample. The lines represent
   predictions from models by \cite{MW02} and
   \cite{2007ApJ...667..760Z} for different minimum masses of dark
   matter haloes (as described above).}
 \label{fig:r0z-nz}
\end{center}
\end{figure*}

Despite the accuracy of our photometric redshifts, as described in
Section~\ref{sec:data-mass-err}, our samples, selected in different
redshift bins, are possibly contaminated by galaxies at other
redshifts. Any contamination will
dilute the original clustering signal, and therefore our measurements
of the amplitude of the correlation function could be underestimated.
From our Monte-Carlo simulations, we estimate the maximum
contamination at a level from 20\% to 30\%, mainly due to galaxies from 
adjacent lower
redshift bins (see Section~\ref{sec:data-mass-err} for a full
description of this).  This
contamination is stronger at higher redshifts. 
However, this effect does not significantly alter our results.
In the worst case, if the contaminating population is uncorrelated, the 
value of $A_{\omega}$ is diluted by a factor of $1/(1-f)^2$ (with $f$ the
contamination fraction), corresponding to an underestimation of our
correlation amplitude by a factor between 1.56 and 2.04 (for 20\% and
30\% contamination by other redshift galaxies).  These factors are
upper limits, and the true value of $A_{\omega}$ is likely to remain
within the error-bars quoted in Table~\ref{tab:results1}.  In any
case, this contamination will only reduce the strength of the
clustering, and again our measures are therefore at worst lower
limits.   This means that any errors in redshifts and/or stellar masses
only dilute our signal, which would make the real clustering strengths
even higher than the ones we observe.  Furthermore such an effect is 
included in our error budget as we are taking the average of measurements 
over six catalogues with perturbed redshift distributions, as described in
Section~\ref{sec:data-mass-err}. This method should account for the
errors due to interlopers.

Moreover, our samples, divided into different stellar mass bins, are
also subject to possible contamination from other stellar mass bins.
Given the mass uncertainties of 0.2-0.3 dex (see
Section~\ref{sec:data-mass-mass}), it is possible that galaxies more
or less massive, and therefore more or less clustered, contaminate our
stellar mass cuts. As mentioned in the Section~\ref{sec:data-mass-err}, 
these effects are more difficult to constrain with our Monte-Carlo 
simulations. However within a given redshift bin, the
measurements of the clustering for samples in adjacent stellar mass
bins are very similar and overlap within their errorbars. The
cross-contamination would therefore only have a minor impact, and in
the sense of increasing the segregation.

\subsection{Spatial correlation lengths and bias}
\label{sec:cf-r0}

In order to compare the clustering of galaxy populations at different
redshifts we derive the spatial correlations for our galaxy samples.
We do this by using the spatial correlation function ($\xi$)
\citep{1977ApJ...217..385G} to derive $r_0$, the co-moving galaxy
correlation length, as defined by:

\begin{equation}
\xi(r,z) = \left( { \frac{r}{r_0(z)} } \right)^{-\gamma}
\label{eq:xir}
\end{equation}

\noindent where $\gamma=1+\delta$. The redshift dependence is included in the
co-moving correlation length $r_0(z)$.  In general the larger the
correlation length ($r_0$), the more clustered the galaxies are.

By measuring the redshift distribution for our samples we can also derive
the correlation length $r_0$ from the amplitude of the angular
correlation $A_{\omega}$ using the relativistic Limber equation
\citep{1999MNRAS.306..988M}. We estimate correlation lengths assuming
that our galaxies are distributed as Top-Hat functions in each of
our narrow redshift bins. Given the derived statistics of our samples
and the narrow redshift bins we are using here, and despite the
accuracy of our photometric redshifts, changing the shape of the
redshift distribution has a negligible impact on the results.

\renewcommand{\arraystretch}{1.2}
\begin{table*}
\begin{center}
 \begin{tabular}{*{9}{c}}
\hline               
z bin$^{(a)}$ & $M_*$ bin$^{(b)}$ &  $<z>$$^{(c)}$ & $<M_*>$$^{(d)}$ & $N^{(e)}$ & $n$$^{(f)}$ & $A_{\omega}$$^{(g)}$ & $r_0$$^{(h)}$ & $b$$^{(i)}$ \\
 & & & & & $\times10^{-4}$ & $\times10^{-3}$ & & \\
\hline               
 0.4--0.8 & 10.0--10.5 & $0.61 \pm 0.12$ & $10.28^{+0.12}_{-0.17}$ & $0.91 \pm 0.17$ & $35.9 \pm 8.1$ & $3.8 \pm 1.9$ & $5.7 \pm 1.5$ & $1.1 \pm 0.3$ \\
 0.8--1.2 & 10.5--11.0 & $0.99 \pm 0.12$ & $10.77^{+0.12}_{-0.17}$ & $1.12 \pm 0.11$ & $25.1 \pm 3.1$ & $3.3 \pm 1.0$ & $5.6 \pm 1.0$ & $1.3 \pm 0.2$ \\
 -- & 11.0--11.5 & $0.98 \pm 0.11$ & $11.19^{+0.12}_{-0.16}$ & $0.53 \pm 0.04$ & $11.9 \pm 1.1$ & $6.6 \pm 2.6$ & $8.2 \pm 1.8$ & $1.8 \pm 0.4$ \\
 -- & 11.0--12.0 & $0.98 \pm 0.11$ & $11.23^{+0.16}_{-0.23}$ & $0.56 \pm 0.04$ & $12.5 \pm 1.0$ & $7.2 \pm 2.2$ & $8.6 \pm 1.4$ & $1.9 \pm 0.3$ \\
 1.2--1.6 & 10.5--11.0 & $1.37 \pm 0.12$ & $10.75^{+0.13}_{-0.18}$ & $0.47 \pm 0.11$ & $8.2 \pm 2.3$ & $6.1 \pm 2.1$ & $7.7 \pm 1.5$ & $2.0 \pm 0.4$ \\
 -- & 11.0--11.5 & $1.38 \pm 0.12$ & $11.20^{+0.12}_{-0.17}$ & $0.23 \pm 0.05$ & $4.0 \pm 1.0$ & $11.8 \pm 4.9$ & $11.1 \pm 2.6$ & $ 2.8 \pm 0.6$ \\
 -- & 11.0--12.0 & $1.38 \pm 0.12$ & $11.25^{+0.16}_{-0.25}$ & $0.25 \pm 0.05$ & $4.3 \pm 1.1$ & $11.5 \pm 5.4$ & $10.9 \pm 3.0$ & $ 2.8 \pm 0.7$ \\
 1.6--2.0 & 11.0--12.0 & $1.73 \pm 0.09$ & $11.28^{+0.18}_{-0.28}$ & $0.11 \pm 0.04$ & $1.7 \pm 0.7$ & $24.0 \pm 18.5$ & $15.2 \pm 6.5$ & $ 4.2 \pm 1.6$ \\
\hline               
\end{tabular}

{\it $^{(a)}$Redshift bin; $^{(b)}$Stellar mass bin in
  Log$(M/M_{\odot})$; $^{(c)}$Mean redshift; $^{(d)}$Mean stellar mass
  in Log$(M/M_{\odot})$; $^{(e)}$Density in arcmin$^{-2}$; $^{(f)}$Abundance in $h^{3}$Mpc$^{-3}$;
  $^{(g)}$Correlation amplitude at 1 degree; $^{(h)}$Correlation
  length in \mpc; $^{(i)}$Bias.}

\caption{Summary of the measurements performed for each sample of galaxies
  selected in redshift and stellar mass. The mean redshift, the mean
  stellar mass, surface density, space abundance, amplitude of clustering at 1 degree
  with a slope of $\delta=0.8$ (according to
  equation~\ref{eq:wtheta}), the correlation length (according to
  equation~\ref{eq:xir}), and the linear bias (according to
  equation~\ref{eq:bias}) are compiled for each sample.}
\label{tab:results1}
\end{center}
\end{table*}
\renewcommand{\arraystretch}{1}

The correlation length is derived from the amplitude of the angular
correlation function in each of our redshift and stellar mass bins,
using each of the six catalogues (the original and the simulated
versions) for each of our three fields. The mean
correlation length is then computed, as well as its associated
error budget, as described in Section~\ref{sec:cf-acf}.
In summary, we find that correlation lengths vary from $5$\mpc to
$15$\mpc for galaxies selected by stellar masses with
$M_{*}>10^{10}$\solmg.  The highest correlation lengths are for the
most massive galaxies at the highest redshifts $z=1.6-2$
(Figure~\ref{fig:r0z-nz}(a)).  In general, at lower stellar masses
and redshifts, the correlation length becomes smaller.

Table~\ref{tab:results1} summarises the values we calculate for
the correlation length $r_0$ measured within our stellar mass
selected samples. In Figure~\ref{fig:r0z-nz} we
plot the correlation length, and the galaxy number density, as a function
of redshift for our stellar-mass-selected galaxy samples. We observe
an apparent decrease in the correlation length with decreasing
redshift, as is also observed in other previous studies
\citep[e.g.][]{2005A&A...439..877L}.  Here the main effect is the
well-known luminosity-segregation, with brighter galaxies (less
abundant) having a larger correlation length
\citep{2006A&A...451..409P,2006ApJ...644..671C}.  As a first rough
approximation, this can be directly linked to a mass-segregation
effect, with more massive galaxies more clustered. Previously,
\cite{2008A&A...479..321M} showed that bright red galaxies, that
are likely very massive, have clustering lengths that are almost
invariant with redshift since $z\sim1$, while less luminous (likely
less massive) galaxies have correlation lengths that decrease with
fainter magnitude at all redshifts.

As shown in Figure~\ref{fig:r0z-nz}(a), we compare our results with
\cite{2008A&A...478..299M}, who have derived correlation lengths from
spatial correlation functions using the VIMOS-VLT Deep Survey (VVDS)
data-sets.  The mean redshift of their sample is
$\overline{z}\simeq0.85$, and for galaxies with masses
Log~$[M_*/(h_{100}^{-2}M_{\odot})])=10.22$ they derived
$r_0=5.06\pm0.49$\mpc, and for galaxies with masses
Log~$[M_*/(h_{100}^{-2}M_{\odot})]=10.70$, $r_0=6.21\pm0.67$\mpc. We
note that these results are in good agreement within the error-bars of
our measurements.

As  analysed in Section~\ref{sec:cf-acf} a contamination of our
different samples by galaxies in adjacent redshift and stellar mass
bins will artificially decrease the measured value of their correlation
length.   This effect is more likely to occur within the higher redshift
samples, and thus the values summarised in Table~\ref{tab:results1}
can be considered as lower limits of the real intrinsic correlation
length.  Such an effect has bin identified in previous analysis as
well \citep[e.g.][]{1999MNRAS.310..540A,2008ApJ...685L...1Q}.  The
net effect of this is that our estimates of the total masses for
these galaxies are also lower limits.
However, even with a contamination fraction at the upper limit of 20\%
and 30\%, as quoted in Section~\ref{sec:cf-acf}, the values of the
correlation lengths we quote in Table~\ref{tab:results1} are only
under-estimated by a factor $\sim1.28$ and $\sim 1.41$ respectively.
The relative uncertainties on our measured correlation length are of
the same order as this correction (Table~\ref{tab:results1}).
This is an upper limit of the correction that could be applied, and
furthermore our correlation length measurements and errors 
take into account these potential effects from interlopers as
described in Sections~\ref{sec:data-mass-err}~and~\ref{sec:cf-acf}.

We also note that \cite{hartley08} have found that the errors on
photometric redshifts can artificially broaden the redshift
distribution which leads to an overestimate of the correlation length
$r_0$. However given the narrow bins we are using in the current
analyses, our redshift distributions are more perturbed by objects
scattered from one redshift bin to another, than from a broadening effect
due to photometric redshift errors. 

In addition, we derive the linear bias of our sample, which relates
the clustering of galaxies to that of the overall dark matter
distribution \citep{2000MNRAS.314..546M}. The rms density fluctuations
of haloes ($\sigma_{8,\rm gal}(z)$) are linked to the rms density
fluctuations of the underlying mass ($\sigma_{8,\rm m}(z)$) by the
bias ($b(z)$) following:

\begin{equation}
b(z)=\frac{\sigma_{8,\rm gal}(z)}{\sigma_{8,\rm m}(z)}
\label{eq:bias}
\end{equation}

\noindent where $\sigma_{8,m}(z)=\sigma_8 D(z)$ is the variance in 
$8$\mpc spheres, assuming that dark matter behaves as predicted
in linear theory \citep[linear growth $D(z)$ -
][]{1992ARA&A..30..499C}, renormalised to the fiducial value
$\sigma_8$ (fixed to $\sigma_8=0.9$ in this study). If the galaxy
correlation function is fit as a power law, then it can be integrated
to give the relative variance in $8$\mpc spheres:
$\sigma_{8,gal}$ \citep{P80}. We find for our sample that the bias
varies from $b\sim1.1$ to $b\sim4.2$ for galaxies selected by stellar
masses $M_{*}>10^{10}$\solmg. The most massive galaxies at the
highest redshifts $z=1.6-2$ are more biased, in good agreement
with biased galaxy formation models \citep{1986ApJ...304...15B}.
Table~\ref{tab:results1} summarises the values of the bias $b$ measured
for our samples.

\renewcommand{\arraystretch}{1.2}
\begin{table*}
\begin{center}
 \begin{tabular}{*{10}{c}}
\hline               
z bin$^{(a)}$ & $M_*$ bin$^{(b)}$ &  $<z>$$^{(c)}$ & $<M_*>$$^{(d)}$ & $M_{\rm min -r_0}$$^{(e)}$ & $f_{r_0}$$^{(f)}$ &$M_{\rm min -n}$$^{(g)}$ & $f_{n}$$^{(h)}$ & $M_{\rm DM}$$^{(i)}$ & $M_*/M_{\rm DM}$$^{(j)}$ \\
 & & & & & & & & & $\times10^{-2}$ \\
\hline               
 0.8--1.2 & 11.0--11.5 & $0.98 \pm 0.11$ & $11.19^{+0.12}_{-0.16}$ & $12.75^{+0.19}_{-0.36}$ & $1.13$ & $12.69^{+0.11}_{-0.15}$ & $0.89$ & $12.70^{+0.21}_{-0.43}$ & $3.15 \pm 2.21$ \\
 -- & 11.0--12.0 & $0.98 \pm 0.11$ & $11.23^{+0.16}_{-0.23}$ & $12.85^{+0.15}_{-0.24}$ & $1.10$ & $12.67^{+0.10}_{-0.13}$ & $0.90$ & $12.80^{+0.17}_{-0.27}$ & $2.63 \pm 1.69$ \\
 1.2--1.6 & 10.5--11.0 & $1.37 \pm 0.12$ & $10.75^{+0.13}_{-0.18}$ & $12.55^{+0.18}_{-0.30}$ & $1.16$ & $12.76^{+0.11}_{-0.16}$ & $0.92$ & $12.49^{+0.19}_{-0.35}$ & $1.85 \pm 1.21$ \\
 -- & 11.0--11.5 & $1.38 \pm 0.12$ & $11.20^{+0.12}_{-0.17}$ & $13.18^{+0.18}_{-0.32}$ & $1.04$ & $12.98^{+0.12}_{-0.17}$ & $0.94$ & $13.17^{+0.19}_{-0.34}$ & $1.08 \pm 0.69$ \\
 -- & 11.0--12.0 & $1.38 \pm 0.12$ & $11.25^{+0.16}_{-0.25}$ & $13.16^{+0.21}_{-0.41}$ & $1.05$ & $12.96^{+0.12}_{-0.18}$ & $0.92$ & $13.14^{+0.21}_{-0.44}$ & $1.29 \pm 1.01$ \\
 1.6--2.0 & 11.0--12.0 & $1.73 \pm 0.09$ & $11.28^{+0.18}_{-0.28}$ & $13.50^{+0.26}_{-0.74}$ & $1.02$ & $13.14^{+0.13}_{-0.20}$ & $0.96$ & $13.50^{+0.26}_{-0.79}$ & $0.61 \pm 0.60$ \\
\hline               
\end{tabular}

{\it $^{(a)}$Redshift bin; $^{(b)}$Stellar mass bin in
  Log$(M/M_{\odot})$; $^{(c)}$Mean redshift; $^{(d)}$Mean stellar mass
  in Log$(M/M_{\odot})$; $^{(e)}$Minimum mass of the Dark Matter Halo
  estimated from the correlation length in Log$[M/(h^{-1}M_{\odot})]$;
  $^{(f)}$Correction factor for accounting of the halo distribution
  effect (estimated from correlation length); $^{(g)}$Minimum mass of
  the Dark Matter Halo estimated from the correlation length in
  Log$[M/(h^{-1}M_{\odot})]$; $^{(h)}$Correction factor for accounting
  of the halo distribution effect (estimated from abundance);
  $^{(i)}$Mass of the Dark Matter Halo as estimated from the
  correlation length and after correction for the halo occupation in
  Log$[M/(h^{-1}M_{\odot})]$; $^{(j)}$Stellar-mass-to-Dark-Matter-mass
  ratio in units of $h$.}

\caption{ Summary of the measurements performed for each sample of galaxies
  selected in redshift and stellar mass. The mean redshift, the mean
  stellar mass, the minimum masses of associated DMHs as inferred by
  the comparison between the DMH model and the correlation length,the
  DMH model and the abundances, and the correction factor to apply to
  take into account the halo distribution effect are compiled for
  each sample. In addition we add the value we use as the mass of the DMH
  and the stellar-mass-to-dark-matter-mass ratio, as inferred by the
  correlation measurements, and corrected for the halo occupation
  effect.}
\label{tab:results2}
\end{center}
\end{table*}
\renewcommand{\arraystretch}{1}

\section{Total dark matter mass of massive galaxies at \MakeLowercase{z} $\mathbf{<2}$}
\label{sec:dmm}

\subsection{Modelling the dark matter halo correlation lengths and abundances}
\label{sec:dmm-cf}
 
The standard CDM model predicts that at any redshift more massive dark
matter haloes are on average more clustered than lower mass systems.  To
model this effect quantitatively, we use the predicted effective bias,
and the abundance evolution, as derived from the \cite{MW02} formalism
for different minimum dark matter halo (DMH) mass thresholds.  This
formalism is based on modelling the effective bias, which relates the
mass fluctuations of haloes, to the mass fluctuations in
spheres that contain an average mass $M$ of underlying dark matter,
according to equation~\ref{eq:bias}.  For this purpose, we use the
halo abundance distribution as a function of mass and redshift
determined by \cite{1999MNRAS.308..119S}, derived from fits to large
N-body simulations, and the linear halo bias for a given mass at a
given redshift, calculated using the function of
\cite{2001MNRAS.323....1S}.  We directly derive the effective bias by
integrating the ratio of these functions on masses above a given
minimum mass $M_{\rm min}$.

A modelling of the correlation length evolution for DMHs of a given
minimum mass $M_{\rm min}$ at a given redshift $z$ can be directly
derived from the effective bias using the recipes described in
\cite{2000MNRAS.314..546M}.  We use also these recipes in the
computation of the linear bias for our sample of galaxies, as
described in Section~\ref{sec:cf-r0}.  Details of this method
can be found in the cited papers.
The predicted correlation lengths and abundances evolutions as a function of 
redshift are shown in Figures~\ref{fig:r0z-nz}(a)~and~(b), for different values of the
minimum mass of DMHs, $M_{\rm min}$.

We note that this model assumes that there is only one galaxy per DMH. Such an
assumption can and often must be incorrect, especially in the case of the
 less massive galaxies in the lowest redshift bins.  If satellite galaxies 
are taken into account, using the same galaxy bias factor measured from the
data, a different $M_{\rm min}$ would be inferred. In order to take into
account the Halo Occupation Distribution (HOD) effect we follow the
simple recipe provided in \cite{2007ApJ...667..760Z}, by summing a
step function for central galaxies and a power law for satellite
galaxies, $\langle N(M)\rangle =1+M/M_1$ for $M>M_{\rm min}$. As
inferred in \cite{2007ApJ...667..760Z}, and following galaxy formation
model predictions \citep[e.g.][]{2005ApJ...633..791Z} and HOD modelling
results \citep[e.g.][]{2005ApJ...630....1Z}, we use $M_1=20 M_{\rm
  min}$.  The modified correlation lengths and abundance evolutions
are shown in Figures~\ref{fig:r0z-nz}(a)~and~(b) with dotted
lines, for different values of the minimum mass $M_{\rm min}$.

\subsection{Dark matter masses of massive galaxies}
\label{sec:dmm-mass}

In Figures~\ref{fig:r0z-nz}(a)~and~(b) we show the predicted relation,
based on dark matter halo models and on Halo Occupation Distribution,
between the correlation lengths and the number densities of dark matter
haloes, and how this relation evolves with redshift.  These
predictions are in good agreement with our observations. As expected,
the mass segregation effect is present, with the correlation lengths
derived for galaxies with higher stellar mass agreeing well with the
predicted correlation lengths of haloes with the highest dark matter
masses.

We go a step further than this and associate the observed correlation
lengths of our stellar mass selected samples with the predicted value
of their host halo masses, thereby providing an estimate of the
typical halo mass for each stellar mass selected bin. To do this, we
make the assumption that only one galaxy is hosted
per halo. Later we apply a correction to take into
account the halo occupation effect.  Finally,  we
investigate the stellar masses of all galaxies within these haloes
to obtain a measurement of the total stellar to halo mass within
dark matter haloes.

This total mass measurement from our clustering analysis gives us a
measure of how massive the haloes that host massive galaxies are. To
obtain these mass measures, we interpolate masses from a grid of
predictions in correlation length/redshift space, from the models
described in Section~\ref{sec:dmm-cf}. Each observed correlation
length at a given redshift is then assigned to a model with a given
DMH minimum mass.  This method has been used previously in papers such
as \cite{2004ApJ...611..685O,2008MNRAS.383.1131M,
  2008ApJ...679..269Y}.  The errors on the correlation lengths are
  propagated to obtain errors on the DMH minimum mass in the same way.
A similar approach is used to estimate the mass of DMHs, by comparing model 
results with the abundances.
  Table~\ref{tab:results2} summarises the values derived for the DMH
  minimum mass $M_{\rm min}$ from their correlation length values and from their abundances.

Our assumption of a single galaxy per halo can result in
overestimating the true value of the DMH minimum mass. Following
\cite{2007ApJ...667..760Z}, we estimate the effect of halo occupation
using a simple prediction described in Section~\ref{sec:dmm-cf}.
However as shown by \cite{2007ApJ...667..760Z}, the effect of the
``one galaxy per halo'' assumption on the true measure of the DMH mass
can be very important at the lower bias levels found in the lower
redshift part of our catalogue. The simple correction we apply
is unlikely to be sufficient to correct for this effect, so we later
discard these sub-samples from our following analysis, as shown in
Table~\ref{tab:results2}.

More detailed models, such as HOD models
\citep[e.g.][]{2001ApJ...554...85W,2005ApJ...630....1Z}, are likely
more appropriate to determine the mass of DMHs associated with a
sample of galaxies.  For instance, in order to have a reliable estimate 
of the mass of the DMHs, such halo occupation models should
fit directly from the angular correlation function determined for the
samples in the lower redshift and stellar masses bins.  However, for our
higher redshift samples with a strong bias, we have estimated our
correlation amplitude on scales $>1$\mpc, where the effects of any extra
halo terms is avoided (see Section~\ref{sec:cf-acf}).
\cite{2004ApJ...608...16Z} have shown that one-halo term clustering is
only observable on scales smaller than this limit. Therefore, our lower mass
galaxies likely reside in the same haloes, but our measurements are
made on scales where galaxies are in separate haloes.  This should
reduce the effect of multiple halo occupation on our estimation of the
DMH mass even for the lower mass samples. One advantage in measuring
clustering over large areas, such as in this study, is the ability to
avoid the one-halo clustering regime, unlike in previous studies using
smaller areas.

As shown in Figures~\ref{fig:r0z-nz}(a)~and~(b), the minimum mass of
dark matter haloes, for a given sample of galaxies, can differ when
estimated from the correlation length measurements or through using
abundance measurements. \cite{2008MNRAS.383.1131M} raised a similar 
issue in their analyses,
and showed that this discrepancy could also come from an erroneous a
priori assumption of the one-parameter halo bias model for a
one-to-one correspondence between dark matter haloes and astrophysical
sources. When we correct for this effect the two estimations are
broadly in agreement. The masses of DMHs derived from the
comparisons between the DMH model (without halo occupation effects
corrected) and the correlation lengths and abundances are summarised
in the Table~\ref{tab:results2}.

\begin{figure}
\begin{center}
  \resizebox{\hsize}{!}{\includegraphics{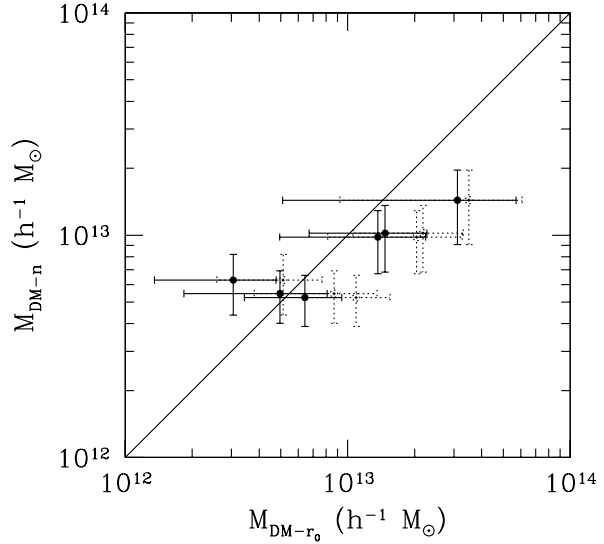}}
  \vspace{-0.5cm}
  \caption[]{Comparison between the masses of DMHs, as estimated using correlation 
    lengths (r$_{0}$) and abundances (n). The masses are corrected from the Halo
    Distribution effects. The plain symbol represent the masses
    estimated from the model using the halo bias function from
    \cite{2001MNRAS.323....1S}, and the dotted symbols from the model
    using the halo bias function from \cite{2005ApJ...631...41T}.}
  \label{fig:mdmr0ab}
\end{center}
\end{figure}

Figure~\ref{fig:mdmr0ab} shows the mass of
DMH inferred by the abundance measurements compared with the one
inferred by the correlation length measurements, with both quantities 
corrected to account for halo distribution effects.
The mild residual discrepancy between the masses estimated by comparing the
models with the correlation length or the abundances, can be explained
by several, likely combined, effects.  

First, and most likely, the contamination effects
between the different bins of our sample could be at the origin of the
discrepancy seen in Figure~\ref{fig:mdmr0ab}. As seen in 
Section~\ref{sec:cf-r0},  the correlation lengths we are using here are 
lower limits, due to possible 
contamination of the samples by interlopers. Therefore the masses of 
the DMHs we measure from the correlation lengths are also lower limits. 
However, this is what we want as the models provide the minimum 
mass of the DMHs, and using the lower-limits is consistent with this. 
Also, interlopers will affect the measurements of the abundances 
of our different samples. 
It is more likely that we are overestimating the number of objects in our less 
populated samples, which are the more massive systems.  In this case, 
the masses of DMHs derived from the abundances of our most massive 
stellar samples are likely underestimated. The combination of these two effects is
likely the origin of the discrepancy between our two estimates, as 
shown in Figure~\ref{fig:mdmr0ab}. These contamination effects are included in 
our error budget as we discuss in Section~\ref{sec:cf-r0}, and, given the 
difficulty in obtaining a precise evaluation of the contamination, we do not correct 
our measurements directly for these effects. 

A second possible explanation, would be incorrect redshift distributions 
for our samples. \cite{2008ApJ...685L...1Q} found a similar
discrepancy between the clustering and the abundance of their colour
selected sample, and invoked an incorrect redshift distribution as a likely origin of 
such discrepancy.   In our case as we are using simple Top-Hat functions,
given the narrow size of our redshift bins, this effect relates
directly to the contamination. 

Furthermore, \cite{2010ApJ...709...67T}
claim that such a discrepancy could originate from cosmic
variance and incorrect models of the halo mass function and halo bias 
function. 
In our case, we take into account cosmic variance in our error
budget, through the field-to-field variance, so we can partially reject 
this explanation.  Finally, our models may be incorrect,
despite being extensively tested. We recompute the model
using a slightly modified halo bias function from the original
function of \cite{2001MNRAS.323....1S}, as proposed by
\cite{2005ApJ...631...41T}. By changing the halo bias function we
only modify the masses estimated from the correlation lengths. As shown
in Figure~\ref{fig:mdmr0ab}, the masses inferred by this new model are
not in better agreement, and the measurements remain almost within
our original errorbars.   We decided to keep the measurements made using
the \cite{2001MNRAS.323....1S} function, as it is more commonly used.
However other inputs in our models, such as the halo mass function, may
also be at the origin of the discrepancy.

\cite{2005MNRAS.363L..66G} have also shown that the clustering of DMHs
depends not only on their halo mass but also on their assembly
history, and especially on their formation time. This ``Halo Assembly
bias'' implies that, at a given mass, DMHs which assembled earlier are
more clustered than DMHs that assembled later. This effect will
increase the discrepancy between masses measured from
the correlation length matching, and from the abundance matching.
\cite{2007MNRAS.374.1303C} have shown that the bias induced by the
environmental dependence of halo formation history at a fixed halo
mass is stronger for faint red central galaxies and weaker for bright
red galaxies.  The assembly bias enhances the two-point correlation by
$\sim 10\%$ for bright galaxies, and suppresses it by the same amount for
fainter galaxies.  However the relative strength in clustering between
our more massive and less massive samples is $\sim 40\%$, much higher
than the effect of the assembly bias itself. Furthermore,
\cite{2005MNRAS.363L..66G} report that this effect is weakest for DMHs
with masses $M_{\rm DM}>10^{13}$\solmm which are within our range.

The masses of the DMHs calculated by matching correlation lengths and
abundances to the models both suffer from 
uncertainties. However as shown in Figure~\ref{fig:mdmr0ab}, the two
estimates are broadly in agreement within our errorbars. We decided
therefore to use the mass estimated from the correlation lengths
corrected from the HOD effect, as we have a more complete error budget
in this case.

\section{The Ratio of stellar and dark matter mass for massive galaxies between $\mathbf{0<}$ \MakeLowercase{z} $\mathbf{<2}$}
\label{sec:ratio}

\subsection{History of mass assembly of galaxies in massive DMHs}
\label{sec:ratio-lit}

\subsubsection{Stellar mass fraction in the central galaxy}
\label{sec:ratio-lit-centr}

\begin{figure*}
\begin{center}
  \resizebox{0.75\hsize}{!}{\includegraphics{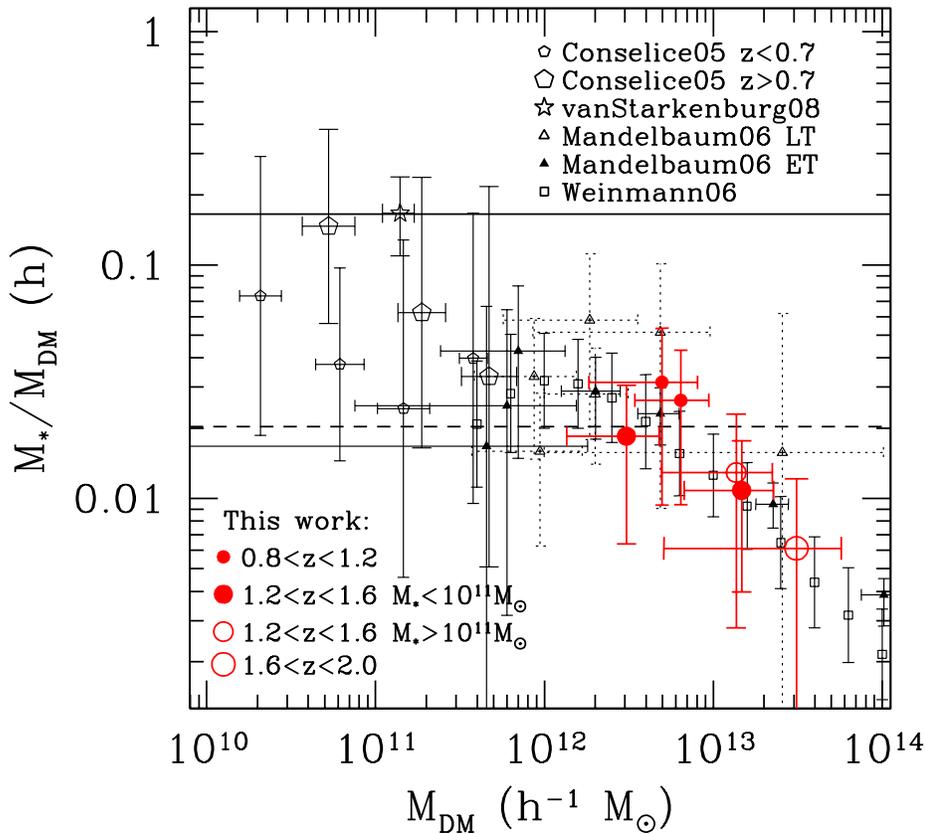}}
  \vspace{-0.5cm}
  \caption[Evolution of the stellar mass fraction with dark matter halo mass 
  and redshift]{Evolution of the stellar mass fraction for massive
    galaxies as a function of the dark matter halo mass ($M_{\rm
      DM}$). The measurements made for this study are shown as red
    open and filled circles, with bigger size symbols corresponding to
    higher redshift samples. Different measurements taken from the
    literature are overplotted, with the size of the symbol bigger for
    samples with $z>0.7$. The continuous line represents the baryonic
    fraction measured from  WMAP5
    \citep{wmap5}, and the dashed line represent the mean
    stellar fraction in the local Universe estimated by
    \cite{2001MNRAS.326..255C}.}
  \label{fig:ratio-lita}
\end{center}
\end{figure*}

\begin{figure*}
\begin{center}
  \resizebox{0.75\hsize}{!}{\includegraphics{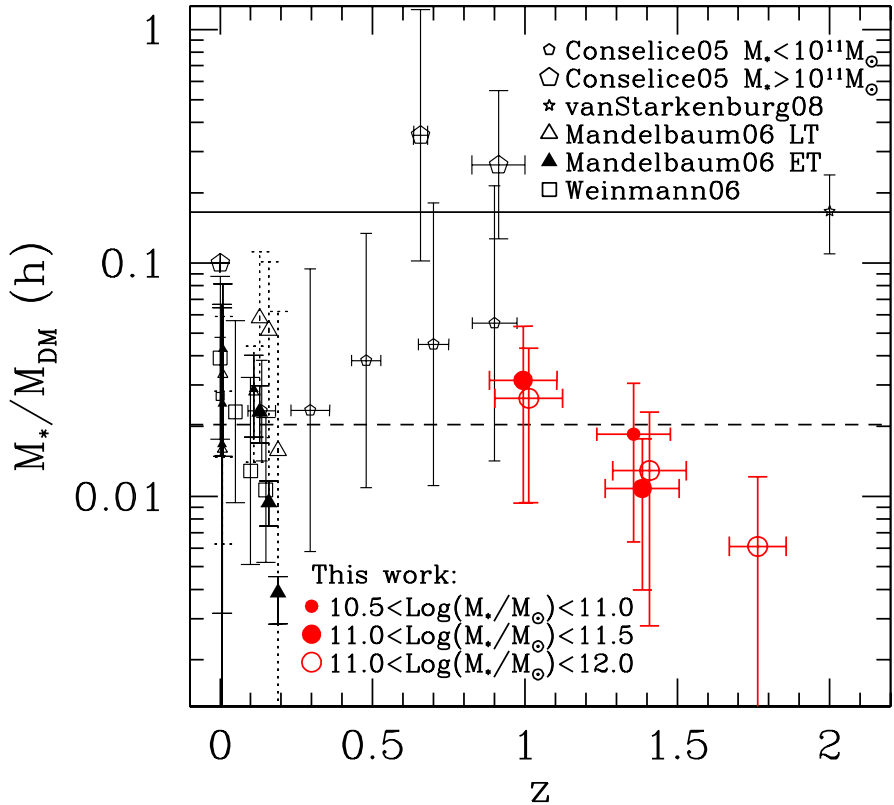}}
   \vspace{-0.5cm}
 \caption[Evolution of the stellar mass fraction with dark matter halo mass 
 as a function of redshift]{Evolution of the stellar mass fraction of massive
   galaxies as a function of redshift. The measurements made in this study are
   shown as red open and filled circles, with bigger sized symbols
   corresponding to more massive samples. Measurements taken from the
   literature are overplotted, with larger symbols for samples with
   $M_*>10^{11}$\solmg. The lines representing the baryonic mass 
  fraction and
   the local stellar mass fraction are the same as described in
   Figure~\ref{fig:ratio-lita}.}
  \label{fig:ratio-litb}
\end{center}
\end{figure*}

Since we know the median stellar mass within each of our sample cuts,
and the clustering and hence total masses of these same galaxies, we 
derive directly the stellar mass fraction as a function of total mass, 
and as a function of redshift for our sample.  We report these values in
Table~\ref{tab:results2}, and display the results on
Figures~\ref{fig:ratio-lita}~and~\ref{fig:ratio-litb}.

In order to study the link between luminous and dark matter we make
several assumptions.  First we assume that the minimum mass of the
DMHs, determined as explained in Section~\ref{sec:dmm-mass}, is an
accurate approximation of the median mass of these DMHs ($M_{\rm
  min}\simeq M_{\rm DM}$). As we are mainly studying DMHs with masses
$M_{\rm DM}>10^{12}$\solmm, and the mass function of DMHs drops
rapidly above this limit, we conclude that this assumption is fair
\citep{2006ApJ...646..881W}.  Our second assumption, for comparison to
other work, is that the virial mass of a system is equivalent to the
total mass in the DMH, which has been confirmed by
\cite{2005MNRAS.363L..11B} using a large N-body simulation. The third
assumption we make, in order to compare our measurements with
literature values, is that the mass in dark matter is equivalent to
the total mass of the system.  Indeed, the \cite{MW02} models we 
use to estimate the mass of the DMHs are based on the behaviour of
dark matter, and do not take into account presence of baryonic matter.
Given that baryonic matter accounts for $\lesssim 15\%$ of the total
density in the Universe \citep{wmap5}, this last assumption is
justified, as $\gtrsim 85\%$ of the density of the system is made up
of dark matter.

As explained in Section~\ref{sec:dmm-mass}, we corrected our halo 
masses for multi-occupation of the haloes. We also focus our analysis
on the highly biased samples, those most likely to have a one-to-one
correspondence between the galaxy and its DMH, sans any outer
satellite galaxies, which is an issue we address later.   We can
therefore assume that we are looking at the stellar mass fraction of
the central galaxies, and are mostly free of any effect due to
satellite galaxies in the present analysis.

To compare our results to previous work, we collected measurements of
stellar and total masses from the literature based on various different
methods to estimate the masses of various DMHs. These include total
masses of disk-like galaxies based on rotation curves at $0.2<z<1.0$,
from \cite{2005ApJ...628..160C}, as well as at $z\sim2.0$, from
\cite{2008A&A...488...99V}. The stellar masses used in these studies
are estimated from SED fits based on spectroscopically confirmed
redshifts, and a wide range of photometry, similar to that described
in Section~\ref{sec:data-mass-mass}, but by using a Salpeter IMF. We
converted all stellar masses into a Chabrier IMF by dividing by a
factor of $\sim 1.5$ \citep{chabrier03}.

We also added measurements of nearby early-type (ET) and late-type
(LT) galaxies from galaxy-galaxy lensing studies in the Sloan Digital
Sky Survey (SDSS) from \cite{2006MNRAS.368..715M}. Stellar masses for
these SDSS galaxies are obtained from the method described in
\cite{2003MNRAS.341...33K}, based on the strength of the
$4000$\AA-break and the Balmer absorption index H$\delta_A$, for a
\cite{kroupa01} IMF (which give estimates very close to the
\cite{chabrier03} IMF). Finally, we use the publicly available
catalogue of groups in the SDSS from \cite{2006MNRAS.366....2W} to
estimate the mass ratio for galaxy groups in the local Universe.
Total masses of these groups are determined by ranking the
completeness-corrected group luminosity, and by assigning to them the
mass of the DMHs, ranked according to their abundances predicted by
models from \cite{MW02}.  The central galaxies of each group are
assumed to be the most luminous galaxy in the group.  
The stellar masses for this sample are determined using the method 
described in \cite{2003MNRAS.341...33K}. 

We add as a continuous line on
Figures~\ref{fig:ratio-lita}~and~\ref{fig:ratio-litb} the value of the
mean baryonic fraction in the Universe, estimated from WMAP5
\citep{wmap5}, which is $f_b=\Omega_b/\Omega_m=0.165\pm0.004$.
From a study of galaxies in the local Universe,
\cite{2001MNRAS.326..255C} measure the stellar mass density to be
$\Omega_* h_{100}=(2.9\pm0.43) \times 10^{-3}$ for a Salpeter IMF. If
we scale this value to a \cite{chabrier03} IMF, and combine it with
the cosmological parameters from WMAP5 \citep{wmap5}, we
derive a mean stellar mass fraction in the Universe of
$f_*=(2.03\pm0.04) \times 10^{-2}$, which is shown in
Figures~\ref{fig:ratio-lita}~and~\ref{fig:ratio-litb} as a dashed
line. The remaining part of the baryonic fraction that is not included
in stars is possibly present in the form of hot and warm gas, as seen
in clusters \citep[e.g.][]{1999MNRAS.305..631A}.

Figure~\ref{fig:ratio-lita} shows that there is a good agreement between 
these very different mass measurements, especially at the massive end.  The
$M_*/M_{\rm DM}$ ratio, even when estimated from completely different
methods, shows a clear decrease at higher halo masses in all
studies. However, estimating accurate DMH masses is
difficult, especially in the less massive cases. Dynamical estimates
(rotation curves, e.g. \cite{2001ARA&A..39..137S}; velocity
dispersions, e.g. \cite{2007MNRAS.374.1169B}) trace the innermost
region of the galaxy, requiring extrapolation to estimate the
total mass of the system.  Weak-lensing requires the use of stacking
methods to provide an accurate estimate of the total mass, and tend to
be difficult to achieve in the case of less massive DMHs
\citep[e.g.][]{2001ApJ...548L...5H,2006MNRAS.368..715M}.  In our case,
galaxies in less massive dark matter haloes are not strongly biased,
making the assumption of a one-to-one correspondence between the
galaxy and the DMH less appropriate, as mentioned earlier.  However
our method, based on clustering properties at large scales, is
efficient enough to determine the luminous-to-dark matter mass ratio
for massive galaxies, and has the advantage of allowing us to estimate
masses at higher redshifts, where for instance lensing cannot be used.

Overall, we find that more massive systems have a lower ratio of
stellar-to-halo-mass, as shown in Figure~\ref{fig:ratio-lita}.  This
behaviour is seen in all of our measurements at $1.0<z<2.0$, as well
as in the local Universe with the SDSS, and thus appears to be
independent of redshift, and is perhaps universal. This relation
implies that, at fixed total mass, the relationship between the mass
of the central galaxy and its DMH does not evolve strongly with cosmic
time.  Physically this may imply that very massive galaxies struggle
to increase their stellar mass in very clustered environments,
resulting in a departure in the luminous-to-dark-matter mass ratio in
massive DMHs compared to lower mass DMHs, which is below the average
value in the local Universe.  These observations also imply that there
is a limit to how much stellar mass a galaxy can have, with a cut off
at a few times $10^{11}$\solmg. This is consistent with the observed
cut-off at high mass of the stellar mass function at low redshift
\citep[e.g.][]{2008MNRAS.388..945B}.

Furthermore, a decreasing stellar mass fraction with increasing halo mass has been
observed in groups and clusters
\citep[e.g.][]{2004MNRAS.355..769E,2004ApJ...617..879L,2007ApJ...666..147G,2007MNRAS.374.1169B}.
\cite{2004ApJ...617..879L} show that Brightest Cluster Galaxies (BCGs)
contribute a large fraction to the total light of their host cluster,
but become progressively less important in the overall luminosity
budget in the highest mass clusters.  This effect is
explained by differential growth between the BCGs and their host
clusters, due to clusters accreting nearby galaxies in lower mass
groups, while the BCGs grow modestly by merging or cannibalism
\citep{2008MNRAS.387.1253W}.  Similarly we can imagine that the
clusters will accrete more dark matter in a similar process while the
central galaxy grows more slowly.

\subsubsection{The stellar mass fraction in satellite galaxies}
\label{sec:ratio-lit-sat}

In Section~\ref{sec:ratio-lit-centr} we focussed our analysis on the
central galaxy of the DMH. Indeed, as explained in
Section~\ref{sec:dmm-mass}, our clustering measurements are made
in a way where we can assume there is only one massive galaxy per
DMH, from which we measure the mass of the host halo. However, an
overabundance of satellite galaxies has been observed in massive DMHs
\citep[e.g.][]{2005MNRAS.356.1233V,2005MNRAS.357..608Y}. It is
therefore possible that a non-negligible fraction of the stellar mass
in the most massive DMHs resides in satellite galaxies
\citep[e.g.][]{2004ApJ...617..879L,2008MNRAS.385.1003B}.

\begin{figure*}
\begin{center}
  \resizebox{0.75\hsize}{!}{\includegraphics{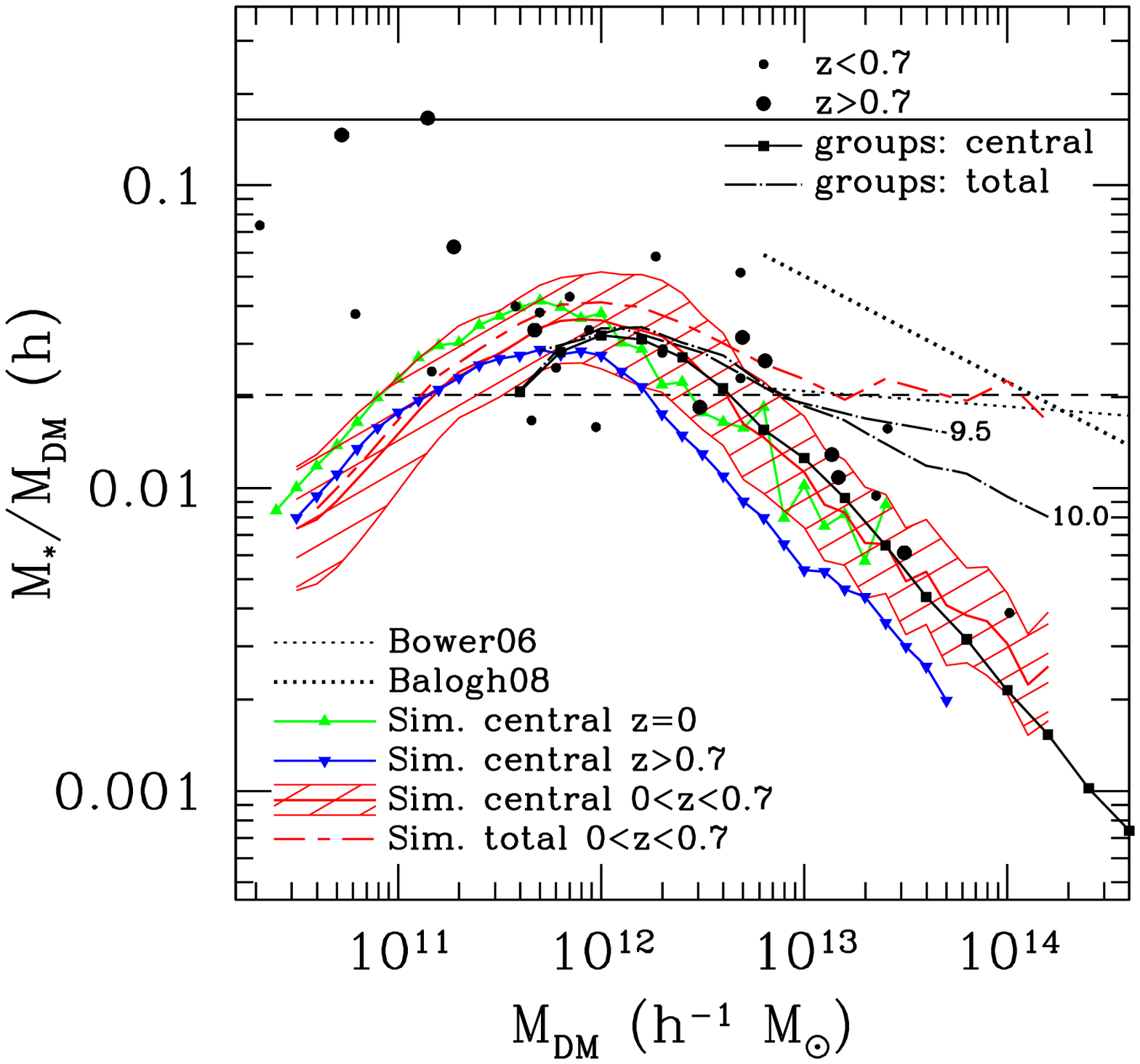}}
  \vspace{-0.5cm}
  \caption[Plot showing the evolution of the stellar mass fraction with dark matter halo mass and redshift]{
Evolution of the stellar mass fraction for massive galaxies as a
    function of dark matter mass ($M_{\rm DM}$). The values for our
    measurements, and previous values from the literature shown in
    Figure~\ref{fig:ratio-lita}, are represented by the dots (small for
    $z<0.7$ and enlarged for $z>0.7$). The full squares connected by a
    full line represent the stellar fraction for the central galaxies
    in SDSS groups, and the long-dash-dotted lines represent the
    stellar fraction for all galaxies in SDSS groups at different
    completeness limits of $M_*=10^{10.0}$\solmg and
    $M_*=10^{9.5}$\solmg, corresponding to redshift limits of $z<0.06$
    and $z<0.045$ respectively.  The two thick straight dotted lines represent
    slopes of the evolution of the total stellar fraction (in central
    and satellite galaxies as well as in the ICL) with DMH masses as
    detailed in \cite{2008MNRAS.385.1003B}, and renormalised
    arbitrarily for more clarity, as we are only interested in
    comparing slopes. The green, red and blue lines represent the
    predictions from the \cite{2007MNRAS.375....2D} model for the
    central galaxies at different redshifts (resp. $z=0$, $0<z<0.7$
    and $z>0.7$). The shaded red region represents the rms scatter of
    the predicted stellar mass fraction for the central galaxies at
    $0<z<0.7$. The red short-dash-long-dash line shows the prediction from the
    model for the total stellar mass fraction estimated from the
    \citeauthor{2007MNRAS.375....2D} model. The horizontal continuous
    line represents the baryonic fraction measured from the WMAP5
    analysis \citep{wmap5}, and the horizontal dashed line
    represents the mean stellar fraction in the local Universe as
    measured by \cite{2001MNRAS.326..255C}.}
  \label{fig:ratio-moda}
\end{center}
\end{figure*}

\begin{figure}
\begin{center}
  \resizebox{\hsize}{!}{\includegraphics{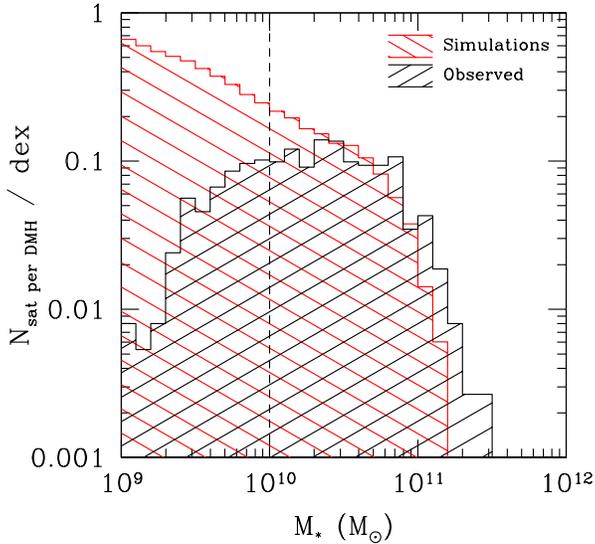}}
  \vspace{-0.5cm}
  \caption{The distribution in stellar mass for satellites in DMHs of masses
    $10^{13.0}$\solmm$<M_{\rm DM}<10^{13.5}$\solmm. The histogram in
    black represents the distribution as observed in SDSS groups at
    $z<0.06$, and the histogram in red is the distribution in the
    Millennium simulation. The vertical dashed line represents the
    completeness limit in stellar mass of the SDSS at $z<0.06$.}
  \label{fig:sat-distr}
\end{center}
\end{figure}

To study this possibility in more detail we use the SDSS group
catalogue \citep{2006MNRAS.366....2W}, as it provides stellar masses
for each member of each group. We limit our analysis of these groups
to those at $z<0.06$ in order to be complete at stellar masses
$M_*>10^{10.0}$\solmg, and at $z<0.045$ for $M_*>10^{9.5}$\solmg. We
plot the ratio of the total stellar mass in the SDSS group DMHs to the
total mass of the DMHs in Figure~\ref{fig:ratio-moda}. As shown in
this figure, the limited completeness of the sample does not allow us
to conclude if the mass in satellite galaxies can account for the
totality of the ``missing'' stellar mass in the DMH.  However, the
stellar mass distribution of satellite galaxies in the SDSS, hosted by
groups with DMHs masses of $10^{13.0}$\solmm$<M_{\rm
  DM}<10^{13.5}$\solmm, is shown in Figure~\ref{fig:sat-distr}. This
study is limited to $z<0.06$, and shows a plateau before reaching the
completeness limit of $M_*>10^{10.0}$\solmg. This flattening at the
faint end is similar to the observations of the luminosity function of
the Local Group \citep[e.g.][]{1999AJ....118..883P}.
  
Furthermore, a decrease in the stellar-mass fraction is also observed in
local clusters. \cite{2006ApJ...640..691V} observed a deficiency
of X-ray gas in galaxy groups compared to clusters.
\cite{2007ApJ...666..147G} explain these observations by the fact that
the missing X-ray gas has cooled in less massive systems to form stars.
This implies that a larger fraction of the baryonic matter in clusters
is in the form of hot gas in more massive systems
\citep[e.g.][]{1999MNRAS.305..631A}.  However, while the total mass
can be derived from the virial mass for these massive clusters, the
total stellar mass is harder to estimate.  A non-negligible
amount of the light emitted by a cluster resides in its Intracluster
Light (ICL), as dynamical processes can strip stars from galaxies
which orbit in the intracluster space \citep[e.g.][and references
therein]{2004ApJ...617..879L,2007ApJ...666..147G}, making the estimate
of the total stellar-mass in a cluster difficult to measure
accurately.

In our study of the stellar-mass fraction from the SDSS group
catalogue, given the likely incompleteness in establishing group
membership, we can only conclude that there is a weak trend in which
the stellar-mass fraction in more massive systems is lower than in
less massive ones, as shown in Figure~\ref{fig:ratio-moda}. Further study 
of the role of satellite galaxies in the total amount of
stellar-mass in a DMH is required to fully understand this result.

\subsubsection{``Halo downsizing'': a possible scenario for the formation of massive DMHs}
\label{sec:ratio-lit-down}

Figure~\ref{fig:ratio-litb} shows that, at fixed stellar mass, the
stellar mass fraction increases at lower redshift.  In other words, the
stellar component apparently grows more rapidly than the dark
component between $1.0<z<2.0$ for galaxies with
$M_*>10^{11.0}$\solmg. However, as shown in Table~\ref{tab:results2},
the average stellar mass in each stellar mass bin does not change
significantly with cosmic time, while the average DMH mass decreases
by a factor $\sim 5$ within these stellar mass selected samples.
Furthermore, the star-formation rate does not change dramatically in
the redshift range $1.0<z<2.0$
\citep[e.g.][]{2005ApJ...630...82P,2007A&A...472..403T,
  2008ApJS..175...48R}, and the high-mass end of the stellar mass function
does not evolve significantly below $z\sim1-2$
\citep[e.g.][]{2005ApJ...619L.131D,2006ApJ...651..120B,2006A&A...459..745F,2007A&A...474..443P,2009ApJ...701.1765M}.
This suggests that, instead of an increase of the average
stellar mass, a statistical process is responsible for an apparent
decrease of the average mass of massive DMHs between $z\sim2.0$ and
$z\sim1.0$ for galaxies with stellar masses $M_*>10^{11.0}$\solmg.

We propose that these observations for very massive galaxies
($M_*>10^{11.0}$\solmg) can be explained by a ``halo downsizing''
effect \citep[e.g.][]{2006MNRAS.372..933N}. We assume that the 
most massive galaxies are already in place at $z\sim2$ in the most 
massive DMHs, in terms of their stellar material. Gradually the stellar 
mass bin we select is populated 
by galaxies that have gained stellar mass with time but that are 
hosted by less massive DMHs than the older systems we find at
higher redshifts. Meanwhile the most 
massive galaxies do not gain in stellar-mass, nor in DMH mass, due to 
a quenching of their star formation and a weak merging rate 
\citep{2008ApJ...680...41D}.  As a
consequence the average mass of the DMHs for these galaxies decreases, and 
the stellar mass ratio increases, which is what we find.

Such an idea is supported by the ``Archaeological downsizing''
\citep{2005ApJ...621..673T}, and the observation that the most massive 
galaxies appear to have assembled their stellar mass earlier than younger 
galaxies. Galaxies with $M_*>10^{11.5}$\solmg assembled half of their 
stellar mass before $z\sim1.5$, and more than 90\% of their mass was 
already in place at $z\sim0.6$ 
\citep{2006ApJ...651..120B, 2008ApJ...675..234P}.  Such
observations imply that the mechanism responsible for
quenching star-formation in galaxies is strongly mass-dependent and
that it occurs earlier in the most massive galaxies
\citep{2006ApJ...651..120B,2008ApJ...681..931B}.  Using
semi-analytical models,
\cite{2008MNRAS.389..567C} have shown that the downsizing effect
observed for red galaxies, could naturally result from a shutdown in
star-formation.  More massive central galaxies in today's Universe
formed earlier and over a shorter period of time than less massive
galaxies. A critical mass of DMHs of $M_{\rm DM}\sim10^{12}$\solmm,
above which a shutdown in star-formation occurs, is also observed in
these studies, corresponding roughly to the mass where we observe a
decline in the stellar-to-dark-matter-mass-ratio in
Figure~\ref{fig:ratio-lita}.

As mentioned in Section~\ref{sec:dmm-mass}, the ``Assembly bias'' 
effect could also mimic the observed increase of stellar-mass ratio
with lower redshifts.  \cite{2007MNRAS.374.1303C} have shown that, at
fixed halo mass, more massive and clustered central galaxies tend to
occupy haloes that formed earlier, while less clustered central
galaxies occupy haloes that formed later. However,
\cite{2007MNRAS.377L...5G} have demonstrated that this effect is
unobservable for massive DMHs ($M_{\rm DM}\sim10^{13}$\solmm).
Although we cannot rule out that assembly bias could artificially
increase the ratio of stellar to dark matter mass which we observe,
this effect is likely not large enough to account for the observed
evolution.

The ``halo downsizing" effect could also be used to explain the 
observation that galaxies in more massive haloes contain a lower
stellar-mass fraction than those in less massive haloes, as seen in
Figure~\ref{fig:ratio-lita}.  A tentative explanation of the formation
of massive galaxies could be that at higher redshift the most massive
DMHs formed rapidly by accretion or mergers
\citep{2003AJ....126.1183C}. This would provide galaxies 
less time to form stars and become more subject to a shutdown of
their star-formation, compared with less massive DMHs
\citep{2006ApJ...651..120B, 2008ApJ...675..234P}.  Moreover, the gas
accreted in the more massive DMHs is hot, which cannot cool quickly in
such massive potentials, suppressing further the star-formation in
these massive galaxies, as observed for groups and
clusters \citep{2007ApJ...666..147G}.

The main difficulties in testing directly the scenario described above
are in measuring accurately the stellar and dark matter masses involved. We
assume within our clustering-derived masses, and through the other
mass estimations, that masses of DMHs, dynamical masses, and total
masses, are equivalent.  However, it is not always easy to compare
dynamical masses with the often used $M_{500}$ and $M_{200}$ masses,
derived for a radius of a system corresponding to overdensities of 500
and 200, relative to the critical density at a given redshift.
Furthermore, dynamical masses are extremely difficult to measure in
low mass systems, particularly at high redshift.  On the other hand,
the total amount of stellar mass and gas mass in the most massive
systems is also difficult to measure, and requires extensive
observations. For instance, a deeper study in the near-infrared bands
of local groups and clusters is required to obtain a better measure of
the stellar mass fractions in the satellites galaxies and in the
ICL. In addition, a better estimate of the gas content of clusters and
groups, through very detailed X-ray studies of a large sample, would
be extremely valuable.

\subsection{Modelling the mass assembly of massive DMHs with semi-analytic models}
\label{sec:ratio-mod}

\subsubsection{Modelling the history of the central galaxy}
\label{sec:ratio-mod-centr}

Semi-analytic models are known to struggle in reproducing the observed
formation of the most massive galaxies, pushing the formation of these
systems to very late epochs
\citep[e.g.][]{1998MNRAS.297L..23K,conselice_mass}. By introducing AGN
feedback
\citep[e.g.][]{2005MNRAS.361..776S,2006MNRAS.370..645B,2006MNRAS.365...11C}
the most recent models now manage to reproduce a star formation
``downsizing'' effect, implying that the stars in more massive
galaxies form early in the Universe \citep{2006MNRAS.366..499D}.  In
order to check if these models are able to reproduce the behaviour we
observe between stellar mass fraction and total mass, how
this fraction evolves with redshift, and to better understand the underlying
physical processes involved, we compare our results to those from one of
the current leading models.  We use the Millennium simulations
\citep{2005Natur.435..629S} in which were implemented the
semi-analytic models developed by \cite{2007MNRAS.375....2D}.

The Millennium model predictions for the stellar mass fraction of the
central galaxy of DMHs, at different redshifts ($z=0$, $0<z<0.7$,
$z>0.7$), are shown in Figure~\ref{fig:ratio-moda}, with the dashed
area showing the rms for predicted DMHs at $0<z<0.7$.  As expected,
the peak in the mass ratio vs.  halo mass occurs at $M_{\rm DM}\sim
10^{12}$\solmm, where star formation is known to be the most
efficient.  In less massive DMHs the infall time of gas clouds is too
long to form stars efficiently, while in more massive haloes it is the
cooling time of the gas cloud that is too long
\citep{1991ApJ...379...52W}.  By construction, the mean value of this
ratio in the models is similar to the observed stellar-mass fraction
in the local Universe. The model from \citeauthor{2007MNRAS.375....2D}
reproduces well the behaviour observed for the central galaxies in the
most massive DMHs at low redshifts. However, at higher redshifts the
agreement is not as good, with the model showing lower
stellar mass fractions compared to our observations. These models seem
to not form massive galaxies quickly enough to reproduce the
observations \citep[see also][for this comparison done on stellar mass
densities]{conselice_mass}.

Previously, \cite{2007MNRAS.376....2K} used
\citeauthor{2007MNRAS.375....2D}'s models to study the high-redshift
galaxy population. They compared the observed mass functions from
\cite{2005ApJ...619L.131D} and \cite{2006A&A...459..745F} with
simulations, and found an overall overprediction by the simulations
between $1<z<3$ in the mass range $10^{10}$\solmg$<M_*<10^{11}$\solmg.
On the other hand their prediction shows a deficiency in this redshift
range for the most massive galaxies with $M*>10^{11}$\solmg.
\cite{2006MNRAS.370..645B} use a different semi-analytic model with a
different recipe for reproducing AGN feedback, also based on the
Millennium simulation, but find results very similar to the
\citeauthor{2007MNRAS.375....2D} model \citep{2007MNRAS.376....2K}.
By comparing the mass functions derived from their models with the
measurements by \cite{2005ApJ...619L.131D},
\citeauthor{2006MNRAS.370..645B} also find an underprediction of the
number densities for the most massive galaxies at high redshifts.
Using the sample of $K$-band selected galaxies at $z\sim0.4-2$
presented here, \cite{conselice_mass} confirm this result by studying
the mass growth of the most massive galaxies. These underpredictions
in the number of massive galaxies at higher redshift by the
semi-analytic models is likely responsible for the disagreement seen
in Figure~\ref{fig:ratio-moda}. On the other hand, observed mass
errors in massive systems can be partially responsible for the
overestimation at the steep end of the mass function
\citep[e.g.][]{2007MNRAS.376....2K}.

Furthermore, \cite{2009ApJ...696..620C} introduced a simple model
based on N-body simulations and abundance matching between galaxies
and DMHs that can shed some light on this problem. This model is tuned
a posteriori to reproduce the observed evolution of the stellar-mass
function and star-formation rate history within galaxies. With this
simple model, \cite{2009ApJ...696..620C} reproduce the shape of the
stellar mass fraction as a function of DMH mass as shown in
Figure~\ref{fig:ratio-moda}.  Their model does not address
explicitly, by construction, the growth history of galaxies more
massive than $M_*>10^{11}$\solmg.  However, their model infers a shift
in the masses of DMH for which the stellar mass fraction peaks from
higher mass at high redshift, to lower mass at lower redshift ($M_{\rm
DM}\sim10^{13}$\solmm at $z\sim2$ to $\sim10^{11.7}$\solmm at
$z\sim0$), where the Millennium simulation predicts the opposite. This
model favours a range of characteristic masses of DMHs for which a
shutdown in star-formation occurs, and is not defined for DMH masses
of $M_{\rm DM}\sim 10^{12}$\solmm \citep[e.g.][]{2008MNRAS.389..567C}.
Our current dataset can not address this issue, but larger and deeper
surveys should be able to observe such a phenomenon.

\subsubsection{Stellar fraction in massive DMHs}
\label{sec:ratio-mod-tot}

As mentioned in Section~\ref{sec:ratio-lit}, a non-negligible amount
of stellar mass within massive dark matter haloes could be contained in
satellite galaxies. We investigate this idea further to determine the
actual stellar to halo mass ratio for our most massive systems.  We
also use the available semi-analytic catalogue derived from
\cite{2007MNRAS.375....2D}, in which central and satellite galaxies
are differentiated, to estimate the total stellar mass in each
simulated DMH.  We do this by adding the stellar mass included in
satellites to the mass of the central galaxies.  We find that in high
mass DMHs the stellar mass fraction almost matches the value of the
mean stellar fraction in the Universe, as shown by the red long-dash-short-dashed
line in Figure~\ref{fig:ratio-moda}. This prediction of a nearly
constant stellar mass fraction is not surprising, as it is an effect
of the construction of the semi-analytical models.  The most massive
systems (groups and clusters) in the models are built from DMHs with a
similar mass distribution, and over a similar time-scale.  They will
therefore form a similar amount of stars, resulting in a similar
stellar-mass fraction.

In Figure~\ref{fig:ratio-moda} we show the predicted evolution of the
stellar mass fraction with mass for DMHs derived by
\cite{2008MNRAS.385.1003B} using the model of
\cite{2006MNRAS.370..645B}, with a nearly horizontal dotted line with
a slope of $d$Log$(M_*/M_{\rm DM})/d$Log$M_{\rm DM}=-0.05$.  The model
from \cite{2006MNRAS.370..645B} leads to similar conclusions as
\cite{2007MNRAS.375....2D}.  Furthermore, the trend found within the
SDSS groups, with a lower mass completeness limit of
$M_*=10^{9.5}$\solmg, seems to converge to the slope determined by the
models.  On the other hand, \cite{2008MNRAS.385.1003B} derived an
estimate of total stellar masses and dynamical masses from the sample
of clusters from \cite{2004ApJ...617..879L} and
\cite{2007ApJ...666..147G}, both taking into account intracluster
light.  The slope derived from their conservative model,
$d$Log$(M_*/M_{\rm DM})/d$Log$M_{\rm DM}=-0.35$, does not match the
stellar mass fraction from the SDSS groups.  This could indicate an
underestimate of the total stellar-mass in their study.

\begin{figure}
\begin{center}
  \resizebox{\hsize}{!}{\includegraphics{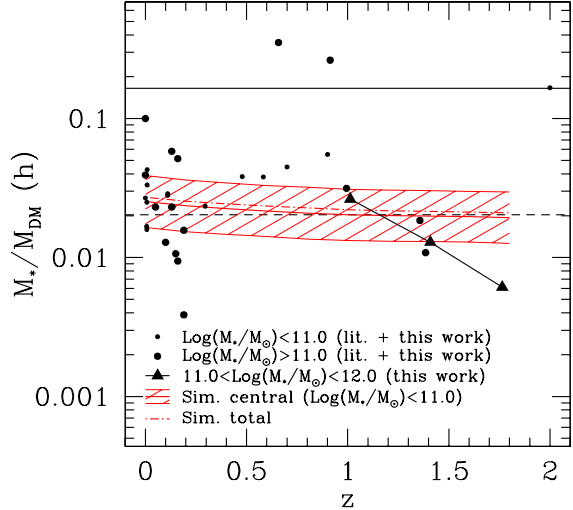}} 
  \vspace{-0.5cm}
  \caption[Plot showing the evolution of the stellar mass fraction with dark matter halo mass and redshift]{
    Evolution of the stellar mass fraction for massive galaxies with
    redshift. The values from our measurements and the literature, as
    shown in Figure~\ref{fig:ratio-litb}, are represented by the dots
    (small for $M_*<10^{11}$\solmg and enlarged for
    $M_*>10^{11}$\solmg). The triangle symbols connected by a
    continuous line represent the measurements made in this study for
    galaxies with stellar masses $10^{11}$\solmg$<M_*<10^{12}$\solmg.
    The red solid line represents the predictions from the
    \cite{2007MNRAS.375....2D} model for the central galaxies at
    $0<z<0.7$ with stellar masses of $M_*<10^{11}$\solmg. The shaded
    red region is the rms scatter of the observed stellar mass
    fractions for the central galaxies at $0<z<0.7$. The red dash-dot
    line shows the prediction from the model for the total stellar
    mass fraction. The horizontal lines represent the baryonic mass
    fraction, and the local stellar mass fraction, as described in
    Figure~\ref{fig:ratio-moda}.}
  \label{fig:ratio-modb}
\end{center}
\end{figure}

In order to verify if the stellar mass fraction derived from the  
SDSS groups, and the predictions from the
\citeauthor{2007MNRAS.375....2D}'s model, are in good agreement, we
show in Figure~\ref{fig:sat-distr} the distribution in stellar mass of
satellite galaxies selected in DMHs of masses
$10^{13.0}$\solmm$<M_{\rm DM}<10^{13.5}$\solmm from the SDSS.  This
figure shows that the satellite galaxies in the SDSS groups reach a
plateau before the completeness limit of $M_*=10^{10}$\solmg, while
the number of satellites in the simulations keeps rising. Such a
flattening at the faint-end is observed in the luminosity function of
the Local Group \citep[e.g.][]{1999AJ....118..883P}, confirming that
simulations fail to reproduce the luminosity/mass function of
satellites galaxies in groups. This is part of the well known missing
satellite problem in CDM \citep[e.g.][]{1999ApJ...524L..19M}.  At the
completeness limits of the SDSS data there are twice as many
satellites per DMH in the simulations than in the observations.
Within the most massive part of this distribution it appears that the
simulations lack massive satellite galaxies when compared to the
observed SDSS groups. This is more likely due to the limited-volume of
the simulations, but could also be due to the lack of massive galaxies
in the overall mass functions as discussed above. However, the
observations contain many uncertainties as well. Estimating the
stellar-masses of faint objects is still tricky, and the group
membership of satellite galaxies, especially at the faint end, is
difficult to ascertain. To solve this issue, a deeper study of groups
and clusters is required to estimate, in a better manner, their total stellar and
dark-matter masses.

Finally, Figure~\ref{fig:ratio-modb} shows the variation of the
stellar-mass fraction with redshift predicted from the simulation
compared to the data themselves.  This figure shows that no variation
of the stellar mass fraction with redshift is predicted by the
simulation for galaxies with stellar masses $M_*<10^{11}$\solmg.
The N-body simulations present a
very tight correlation between the mass of the DMHs and the number of
galaxies hosted by each DMH at any epoch, and are in this sense very
close to Smooth Particle Hydrodynamic simulations and analytic Halo
Occupation models
\citep[e.g.][]{2003ApJ...593....1B,2005ApJ...633..791Z,2008ApJ...678....6W}.
Therefore, the stellar mass fraction in a given range of stellar mass
is constant with epoch, and by construction equal to the mean value in
the local Universe, explaining their failure to reproduce the observed
increase in stellar mass fraction with lower redshift.  Moreover, as
shown in Figure~\ref{fig:ratio-modb}, the effect of the fraction of
mass in satellites is negligible here as well.  However, because of
the limited volume of the simulations, the number of galaxies with
$M_*>10^{11}$\solmg does not allow for a direct comparisons with our
observations. Simulations over larger volume are required, and
observations of the stellar mass ratio of less massive galaxies at
$1.0<z<2.0$ would be very useful as well. 

\section{Summary}
\label{sec:summ}

We present in this paper an analysis of the clustering properties of a
stellar mass selected sample of galaxies over 0.7 deg$^{2}$ taken from
the Palomar Observatory Wide-Field Infrared Survey (POWIR).  By
utilising information from three separate fields, we measure the
two-point correlation function, and based on this we derive the
spatial correlation length and bias for our sample of galaxies within
various stellar mass bins at $M_{*}>10^{10}$\solmg.  Our major results
are:

\vspace{0.2cm}

\noindent I. We find that the correlation length ($r_{0}$) varies with both
redshift and stellar mass. We find that the largest correlation
lengths (most clustered systems) are found for galaxies at the highest
stellar masses with $10^{11}$\solmg$<M_{*}<10^{12}$\solmg at $z\sim2$.
The correlation length for galaxies at all stellar mass selections are
found to decrease at lower redshift.

\vspace{0.2cm}

\noindent II. We derive the dark matter halo masses for these stellar mass
selected systems utilising the correlation lengths and models from
\cite{MW02}.  Due to a likely contamination between our samples, the correlation
lengths and therefore the  dark matter halo masses are lower limit estimates. However
we take into account this effect in our error budgets. We find that halo masses for our 
galaxies selected by $M_{*}>10^{10.5}$\solmg are a factor of 30 to 170 times the stellar
masses of each galaxy.  This is in very good agreement with other
methods for measuring halo masses, including lensing and kinematics.

\vspace{0.2cm}

\noindent III. We find a remarkable relation between dark matter
halo mass ($M_{\rm DM}$), and the ratio of stellar-to-dark-matter-mass
($M_{*}/M_{\rm DM}$) for the central galaxy in the halo, and find
that this relation does not vary much with redshift.  This correlation is
such that the more massive a dark matter halo mass $M_{\rm DM}$ is,
the lower the ratio of stellar-to-halo-mass ($M_{*}/M_{\rm DM}$).
We further find that this correlation exists over at least two orders of
magnitude from halo masses $M_{\rm DM}=10^{12.0}$ to $10^{14.0}$\solmm.
This is true for all methods of determining halo masses, from lensing,
kinematics, and clustering.

\vspace{0.2cm}

\noindent IV. Within our massive galaxy sample at $z\sim1-2$ we find
that the ratio of stellar-to-halo-mass increases from high to low
redshifts.  This correlation implies that at a given stellar mass
selection, the average underlying mass of the host halo decreases at lower
redshifts.  This may imply that the most stellar massive systems at
the highest redshifts are hosted by very massive haloes, but at lower
redshift they stop forming new stars, and therefore do not increase their
stellar mass, while other systems in the lower mass dark matter
haloes grow to reach similar stellar mass.  This is an example of
``Halo downsizing''. The `Halo Assembly bias' implying that clustering
depends not only on the mass of the dark matter halo, but also on the
assembly history and environment of the galaxy, can also contribute
partially to this observed increase in the stellar mass fraction with
redshift.

\vspace{0.2cm}

\noindent V. We compare our results to the semi-analytic models built using the
Millennium simulation.  We find that there is roughly a good agreement
between models and the data for the central galaxy. The predicted 
value of the total-stellar-to-total-halo mass ratio
decreases only slightly with increasing halo mass, also in rough
agreement with our observational results.  We use the SDSS group
catalogue from \cite{2006MNRAS.366....2W} to examine whether the
missing stellar mass can be accounted for by observed satellite
galaxies. We find that, even though the overall agreement is good, the
model marginally overpredicts the number of faint galaxies in
comparison to the SDSS catalogue. Moreover the Millennium simulation
fails to reproduce the ``halo downsizing'' effect we observe. We argue
that this missing mass, in the more massive haloes, is most probably
in the form of baryons in a warm/hot phase.

\section*{Acknowledgements}
We thank the anonymous referee for his detailed and
constructive report.  

SF, CJC, WGH and SPB  acknowledge 
support from the Science and Technology Facilities Council (STFC).

The Palomar and DEEP2 surveys would not have been completed without
the active help of the staff at the Palomar and Keck observatories, and 
the DEEP2 and POWIR teams, particularly Sandy Faber and Richard Ellis.
Funding to support this effort came from a National Science Foundation
Astronomy \& Astrophysics Fellowship, and grants from the UK Particle
Physics and Astronomy Research Council (PPARC).

We also thank Simone Weinmann and Xiaohu Yang for making
their group catalogue publicly available. We are also grateful to the
SDSS collaboration for producing a wonderful data set and for making
their products publicly available.

The Millennium Simulation databases used in this paper and the web
application providing online access to them were constructed as part
of the activities of the German Astrophysical Virtual Observatory.

\bibliographystyle{aa}


\label{lastpage}

\end{document}